\documentclass[a4paper,11pt]{article}
\pdfoutput=1 

\usepackage{jinstpub} 

\usepackage{subcaption}
\usepackage{lineno}

\title{\boldmath Embedded Readout Electronics
       R\&D for the Large PMTs in the JUNO Experiment}

\author[a]{M.~Bellato,}
\author[a]{A.~Bergnoli,}
\author[a,b]{A.~Brugnera,}
\author[c]{S.~Chen,}
\author[d]{Z.~Chen,}
\author[e]{B.~Clerbaux}
\author[a]{F.~dal~Corso,}
\author[a]{D.~Corti,}
\author[c]{J.~Dong,}
\author[b]{G.~Galet,}
\author[a,b]{A.~Garfagnini\note{Corresponding author: alberto.garfagnini@pd.infn.it},}
\author[a,b]{A.~Giaz,}
\author[c]{G.~Gong,}
\author[f]{C.~Grewing,}
\author[d]{J.~Hu,}
\author[a]{R.~Isocrate,}
\author[d]{X.~Jiang,}
\author[c]{F.~Li,}
\author[a]{I.~Lippi,}
\author[a,b]{F.~Marini,}
\author[d]{Z.~Ning,}
\author[g]{A.~G.~Olshevskiyi,}
\author[a]{D.~Pedretti,}
\author[e]{P.A.~Petitjean,}
\author[f]{M.~Robens,}
\author[g]{V.~Shutov,}
\author[h]{A.~Stahl,}
\author[h]{J.~Steinmann}
\author[d]{Y.~Sun,}
\author[f]{S.~van~Waasen,}
\author[d]{Y.~Wang,}
\author[d]{Z.~Wang,}
\author[d]{W.~Wei,}
\author[d]{X.~Yan,}
\author[e]{Y.~Yang,}
\author[i]{A.~Aiello,}
\author[i]{A.~Andronico,}
\author[j]{V.~Antonelli,}
\author[k]{W.~Bandini,}
\author[j]{A.~Brigatti,}
\author[m,n]{A.~Barresi,}
\author[o]{A.~Budano,}
\author[i]{R.~Bruno,}
\author[p]{A.~Cabrera,}
\author[m,q]{A.~Cammi,}
\author[i,l]{R.~Caruso,}
\author[m,n]{D.~Chiesa,}
\author[r]{C.~Clementi,}
\author[i,l]{S.~Costa,}
\author[j,s]{X.~Ding,}
\author[a]{S.~Dusini,}
\author[o]{A.~Fabbri,}
\author[i,l]{M.~Fargetta,}
\author[k]{G.~Fiorentini,}
\author[j,t]{R.~Ford,}
\author[j]{A.~Formozov,}
\author[j]{M.~Giammarchi,}
\author[a,b]{M.~Grassi,}
\author[j]{C.~Landini,}
\author[j]{P.~Lombardi,}
\author[i,l]{C.~Lombardo,}
\author[g]{Y.~Malyshkin,}
\author[k]{F.~Mantovani,}
\author[o]{S.~M.~Mari,}
\author[o]{C.~Martellini,}
\author[u]{A.~Martini,}
\author[j]{E.~Meroni,}
\author[a]{M.~Mezzetto,}
\author[j]{L.~Miramonti,}
\author[o]{P.~Montini,}
\author[k]{M.~Montuschi,}
\author[m,n]{M.~Nastasi,}
\author[r]{F.~Ortica,}
\author[u]{A.~Paoloni,}
\author[j]{S.~Parmeggiano,}
\author[r]{N.~Pelliccia,}
\author[m,n]{E.~Previtali,}
\author[j]{G.~Ranucci,}
\author[o]{D.~Riondino,}
\author[j]{A.~C.~Re,}
\author[k]{B.~Ricci,}
\author[r]{A.~Romani,}
\author[j]{P.~Saggese,}
\author[o]{G.~Salamanna,}
\author[a,b]{F.~H.~Sawy,}
\author[k]{A.~Serafini,}
\author[o]{G.~Settanta,}
\author[a,b]{C.~Sirignano,}
\author[m,n]{M.~Sisti,}
\author[a]{L.~Stanco,}
\author[k]{V.~Strati,}
\author[i,l]{C.~Tuv\'e,}
\author[i,l]{G.~Verde,}
\author[u]{L.~Votano,}

\affiliation[a]{INFN Sezione di Padova, Padova, Italy}
\affiliation[b]{Universit\`a di Padova, Dipartimento di Fisica e Astronomia, Padova, Italy}
\affiliation[c]{Tsinghua University, Beijing, China}
\affiliation[d]{Institute of High Energy Physics, Beijing, China}
\affiliation[e]{Universit\`e Libre de Bruxelles, Brussels, Belgium}
\affiliation[f]{Forschungszentrum J\"ulich GmbH, Central Institute of Engineering, Electronics and Analytics - Electronic Systems(ZEA-2), J\"lich, Germany}
\affiliation[g]{Joint Institute for Nuclear Research , Dubna, Russia}
\affiliation[h]{III Physikalisches Institut B, RWTH Aachen University, Aachen, Germany}
\affiliation[i]{INFN Sezione di Catania, Catania, Italy}
\affiliation[j]{INFN Sezione di Milano e Universit\`a di Milano, Dipartimento di Fisica, Milano, Italy}
\affiliation[k]{INFN Sezione di Ferrara e Universit\`a di Ferrara, Dipartimento di Fisica e Scienze della Terra, Italy}
\affiliation[l]{Universit\`a di Catania, Dipartimento di Fisica e Astronomia, Catania, Italy}
\affiliation[m]{INFN Sezione di Milano Bicocca, Milano, Italy}
\affiliation[n]{Universit\`a di Milano Bicocca, Dipartimento di Fisica, Milano, Italy}
\affiliation[o]{INFN Sezione di Roma Tre e Universit\`a di Roma Tre, Dipartimento di Matematica e Fisica, Roma, Italy}
\affiliation[p]{IJC Laboratory, CNRS/IN2P3, Université Paris-Saclay. 91405 Orsay. France}
\affiliation[q]{Politecnico di Milano, Dipartimento di Energetica, Milano, Italy}
\affiliation[r]{INFN Sezione di Perugia e Universit\`a di Perugia, Dipartimento di Chimica, Biologia e Biotecnologie, Perugia, Italy}
\affiliation[s]{Gran Sasso Science Insitute, L'Aquila, Italy}
\affiliation[t]{SNOLAB, Lively, Ontario, Canada}
\affiliation[u]{INFN Laboratori Nazionali di Frascati, Italy}

\abstract{
Jiangmen Underground neutrino Observatory (JUNO)
is a next generation liquid scintillator neutrino experiment under
construction phase in South China.  Thanks to the anti-neutrinos
produced by the nearby nuclear power plants, JUNO will primarly
study the neutrino mass hierarchy, one of the open key questions in neutrino physics.
One key ingredient for the success of the measurement is to use high speed,
high resolution sampling electronics located very close to the detector
signal. Linearity in the response of the electronics is another important
ingredient for the success of the experiment.
During the initial design phase of the electronics, a custom
design, with the Front-End and Read-Out electronics located very close to
the detector analog signal has been developed and successfully tested.
The present paper describes the electronics structure and the first tests
performed on the prototypes. The electronics prototypes have been tested
and they show good linearity response, with a maximum deviation of
1.3\% over the full dynamic range (1-1000 p.e.), fullfilling the JUNO
experiment requirements.}

\keywords{Only keywords from JINST's keywords list please}

\arxivnumber{1234.56789} 

\begin{document}
\maketitle
\flushbottom

\section{Introduction}
\label{sec:intro}
The Jiangmen Underground Neutrino Observatory (JUNO)~\cite{bib:juno:yb}
is a next generation
neutrino experiment under construction in South China.
Thanks to the nearby Yangjiang and Taishan nuclear power plants, JUNO
will attack the open question of neutrino mass hierarchy by measuring
the inverse beta decay interactions of reactor anti-neutrinos in
the detector.
The JUNO detector structure~\cite{bib:juno:cd} is quite simple but
impressive: a large acrylic sphere (34.5~m diameter),  kept in position
by a stainless steel truss, contains almost 20 kton of ultra pure
liquid scintillator - Linear Alkyl Benzene as solvent, with the scintillating
PPO fluorine (2.5-Diphenyloxazole) and a wavelength shifter (bis-MSB) diluted.
The stainless steel support structure holds the inner vessel and
almost 20000 large (20-inch) PMTs and about 25000 small (3-inch)
PMTs~\cite{bib:double-calorimetry}.
The described central detector will be placed inside an
instrumented water pool that will act both as a Cherenkov muon veto and as
shield against environmental radiation (gammas and neutrons) coming
from the rock.
Finally, a top tracker made with the plastic scintillator
detectors of the former OPERA~\cite{bib:opera:det}
experiment at Gran Sasso~\cite{bib:opera:tt} will be placed on top of the
water pool.

A key ingredient for the measurement of the neutrino mass hierarchy
is an excellent but challenging energy resolution of the central detector:
3\% at 1 MeV or better is required.
Moreover, independently of the energy resolution and thanks to the large
statistics, JUNO is going to precisely measure the neutrino mixing parameters,
$\theta_{12}$, $\Delta m_{21}^2$ and $\Delta m_{ee}^2$ with an ultimate
sensitivity, below the 1\% level~\cite{bib:juno:yb}.
Beyond mass hierarchy and precision determination of the neutrino
oscillation parameters, a large liquid scintillator detector can give access to
valuable data on many topics in astroparticle physics, like
supernova burst and diffuse supernova neutrinos, solar neutrinos, atmospheric
neutrinos, geo-neutrinos, nucleon decay, indirect dark matter searches and a
number of additional exotic searches. A reference to the JUNO rich physics
program can be found elsewhere~\cite{bib:juno:yb}.

The Front-End and Read-Out electronics for the large PMTs system are
an important component and their performance is crucial for the
successes of the measurements. This translates in a very good resolution
both in single photon detection and in multi photon signal.
The overall requirements coming from physics are the following:
\begin{itemize}
\item signal range: from 1 p.e. to 100 p.e. with a linear response and
charge resolution from 0.1 p.e. to 1 p.e.; this requires that the noise level
must remain below 0.1 p.e. for single p.e. detection.
\item background range: from 100 p.e. to 1000 p.e. with a resolution
of 1 p.e.
\item signal rise time around 2.5~ns. The requirement translates in a
bandwidth of about 400~MHz and therefore a sampling rate of 1Gsample/s
is appropriate.
\end{itemize}

To achieve such goals, considering the detector structure and topology,
the Read-Out electronics has to be positioned very close to the PMTs.
This novel concept, compared to legacy large scintillator based neutrino
experiments, (see for instance~\cite{bib:borex} and~\cite{bib:kamland}),
allows to reach the best performances in terms of signal to noise
ratio since the analog part of the signal is digitized at a very early stage;
moreover the data readout throughput is lowered thanks to the reduced number
of cables needed to communicate to the back-end electronics and since local
data storage is possible, it opens the possibility to perform complex
signal pre-processing tasks locally, before data is sent to the DAQ.
On top of that, several constraints affect the electronics
design: as an example it must satisfy high reliability criteria since
it can't be repaired or replaced in case of malfunctions or breakdown.
Moreover it has to be designed with low power consumption in mind to
minimize the single channel power consumption and fit in a limited space
requirement for the installation.

According to~\cite{bib:juno:cd} the following guidelines have been
identified for the electronics design:
\begin{itemize}
\item[-] positioning of the Front-End and Read-Out electronics
         close to the PMT output signal;
\item[-] usage of high speed and high resolution waveform digitizers
         with large bandwidth;
\item[-] exploitation of signal processing and local data storage,
         very close to the PMT;
\item[-] interface to the DAQ and Trigger electronics through Ethernet
         cables;
\item[-] Power over Ethernet and synchronous signals transport (Clock and
         Trigger) through the same Ethernet cable;
\item[-] single channel power consumption not greater than 10~W;
\item[-] high reliability of the PMT electronics: less than 0.5\% of
         malfunctioning or broken channels in six years of data taking.
\end{itemize}

The present paper reports on the result of an R\&D effort carried on
inside the JUNO collaboration to design and test an electronics
readout scheme as a possible candidate for the final large PMT electronics.

\section{The electronics scheme}
\label{sec:BX}
\begin{figure}[h]
\centering
\includegraphics[width = 0.95\textwidth]{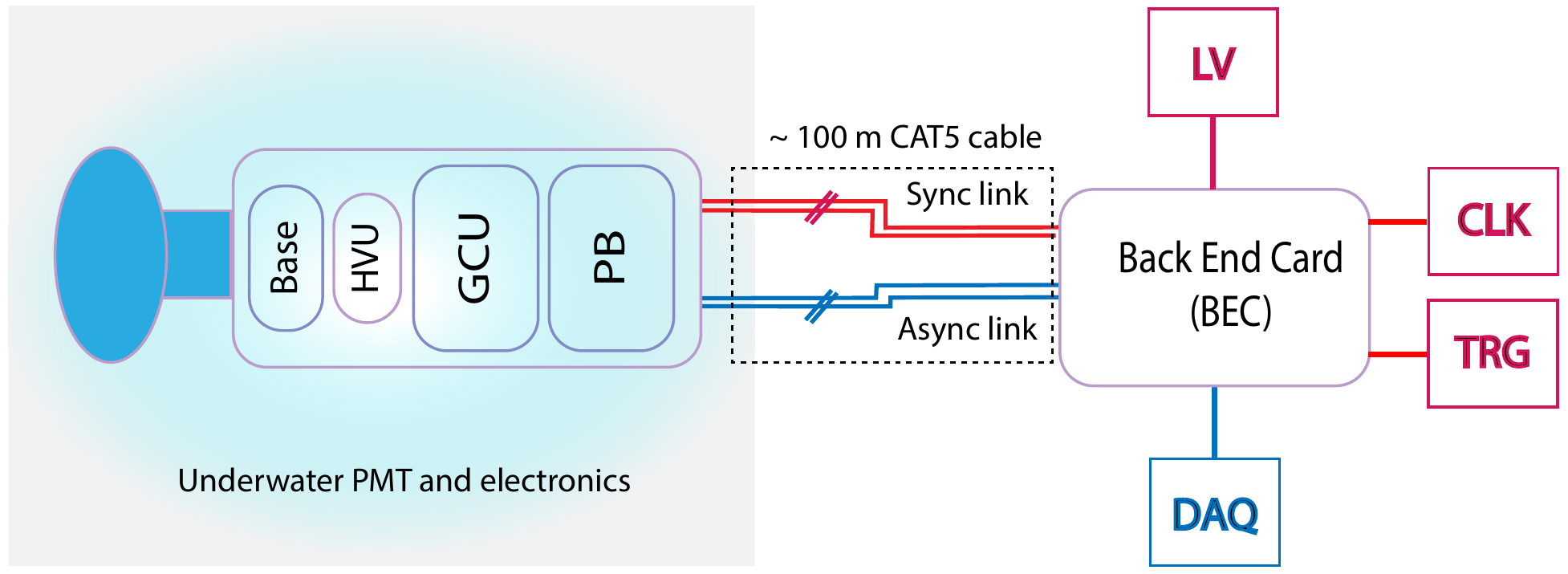}
\caption{Electronics scheme of the JUNO large PMT electronics.
The 'wet' electronics (left) is connected to the 'dry' electronics (right)
by means of a 100~m long CAT5 cable (middle).}
\label{fig:BX:scheme}
\end{figure}

The structure of the electronics is shown in Figure~\ref{fig:BX:scheme}.

The electronics is split into two parts: one located on the PMT,
in the underwater water tank and henceforth referred
to as 'wet' electronics, and the 'dry' electronics in the electronics room.

The 'wet' electronics is made of the following components
(see Figure~\ref{fig:BX:scheme}, from left to right):
\begin{itemize}
\item[-] Base: the PMT voltage divider and splitter;
\item[-] High Voltage Unit (HVU): a programmable module which provides
         the bias voltage to the voltage divider;
\item[-] Global Control Unit (GCU): the intelligent part of the
         'wet' electronics.
         It receives the analog signal, digitizes it and processes the digital
         output.
\item[-] Power and Communication Board (PB): the interface to the 'dry'
         electronics. It drives the synchronous Clock (CLK) and Trigger (TRG)
         links and provides power to the 'wet' components.
\end{itemize}

The 'dry' electronics is composed of the Back End Card (BEC)
which receives and sends the digital data, distributes the
power, and handles the synchronous signals (CLK and TRG),
the trigger electronics, which will be described elsewhere,
the central JUNO clock synchronized to GPS, and the power supplies.

The communication between the dry and wet parts uses a standard CAT5e cable.
The four twisted pairs of the cable accommodate:
\begin{itemize}
\item[-] an asynchronous down-link using the 100BASE-TX fast ethernet
         communication standard.
\item[-] an asynchronous up-link using the same 100BASE-TX fast ethernet
         standard. 
\item[-] a synchronous 62.5~MHz clock signal which is
         derived from the central JUNO clock.
\item[-] a Trigger input sending a digital "{\bf{1}}" every 16 ns,
         if a photon is detected by the PMT. The trigger decision, initiating
         the readout of the PMT, is distributed through the asynchronous
         down-link.
\end{itemize}
Power is transmitted through a static voltage difference between the two
wires of the twisted pairs. The digital power is transmitted on both
asynchronous links at a voltage of 24~V; the analog power uses the clock
line with the same voltage. 

\begin{figure}
\begin{minipage}[c]{1.1\textwidth}%
\hspace{-1.0cm}%
\includegraphics[scale=0.325]{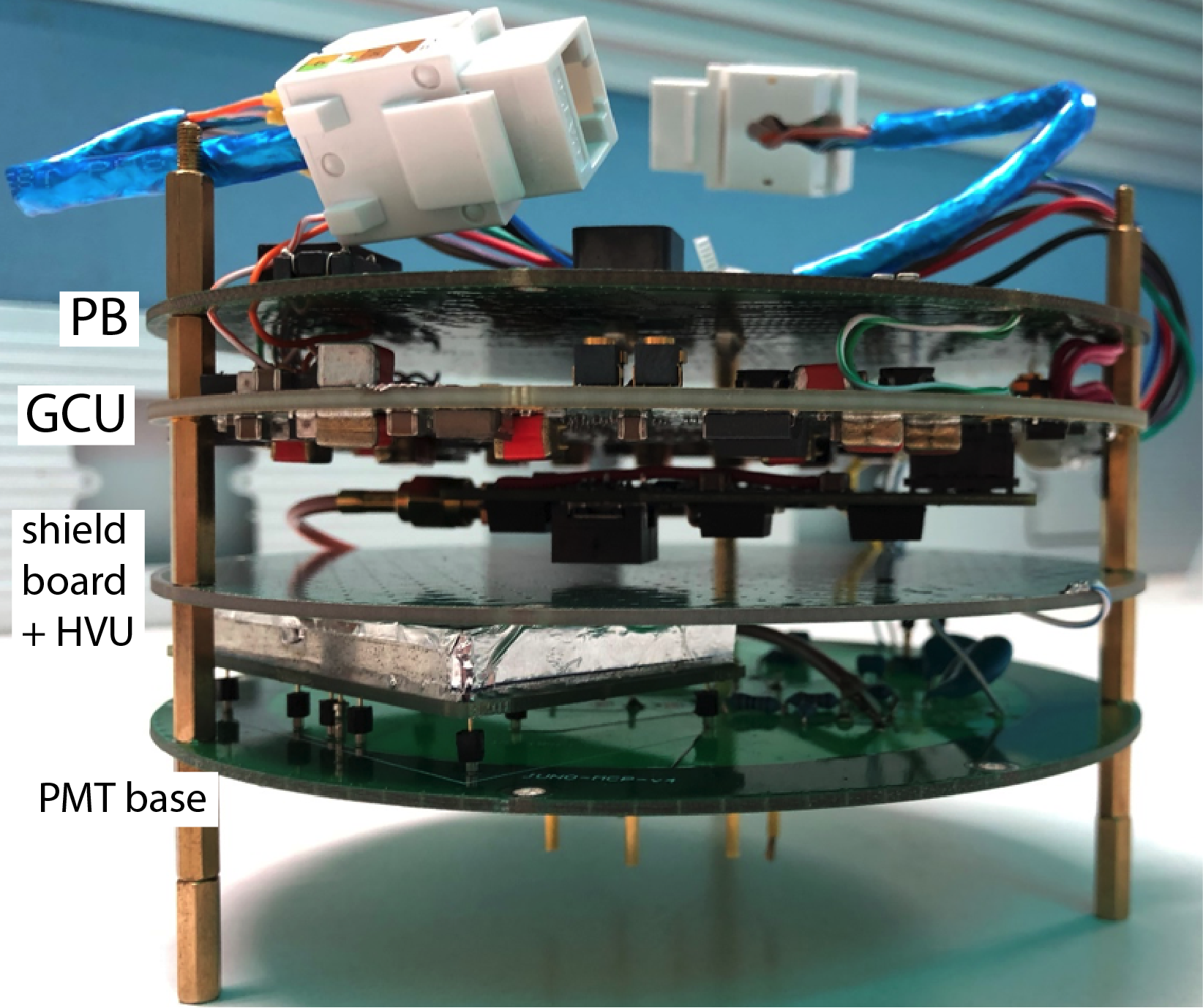}
\includegraphics[scale=0.4]{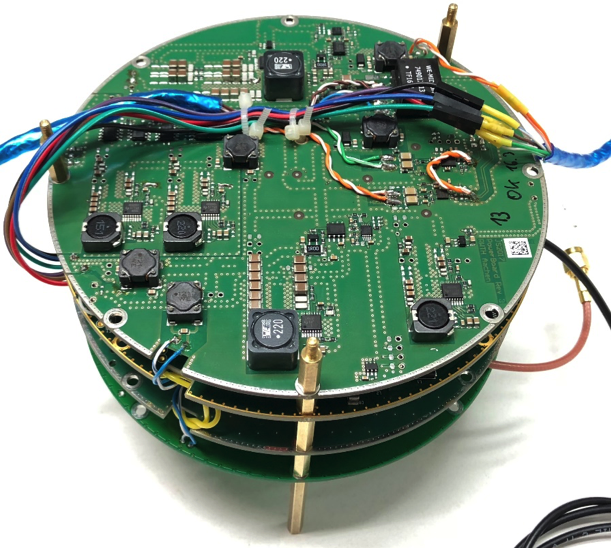}
\end{minipage}%
\caption{Assembly of the prototype boards:
lateral (left) and top (right) views.}
\label{fig:proto:bx_castle}
\end{figure}

A realization of the prototype boards
assembled in a castle-like configuration, before being coupled to the PMT,
is shown in Figure~\ref{fig:proto:bx_castle}.
In the lateral view (Figure~\ref{fig:proto:bx_castle}, left),
from top to bottom the following boards can be seen: PB,
GCU, an empty shielding board, and the PMT base.
The HVU is on top of the PMT base, touching the shielding board.
The diameter of the boards is about 140~mm, while
the height of the assembly is about 100~mm. The Ethernet
socket, which was used to test the prototypes, is visible on the top
of Figure~\ref{fig:proto:bx_castle}, left.

A full view of the PB, with all the components, is available on the right
plot of Figure~\ref{fig:proto:bx_castle}.
Connections between the different boards are made with cables soldered on
to the PCBs.

In the following sections, a description of the different boards is given.

\subsection{PMT voltage divider and High Voltage Unit}
\label{sec:BX:base}
\begin{figure}[htbp]
\centering
    \includegraphics[width=0.8\columnwidth]{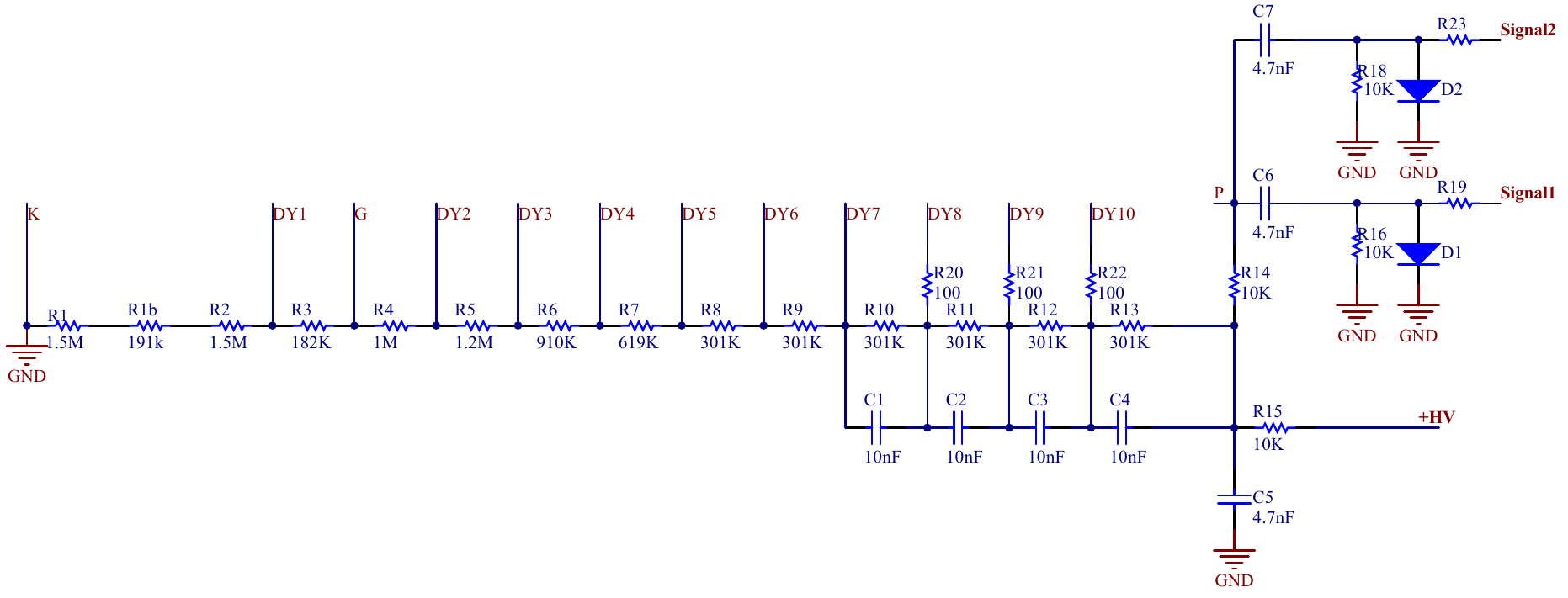}
    \includegraphics[width=0.8\columnwidth]{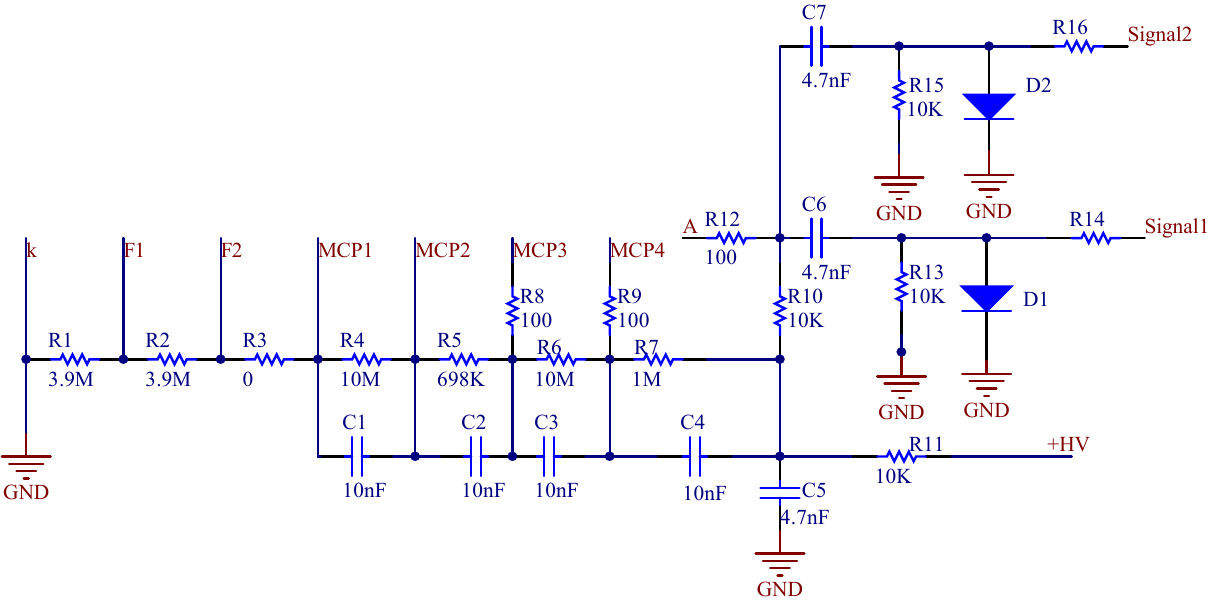}
  \caption{\label{fig:BASE_schematics}Top: Hamamatsu PMT voltage divider
           schematics. Bottom: MCP-PMT voltage divider schematics.}
\end{figure}

JUNO will deploy, in total, about 20000 large size PMTs
 of two different types~\cite{bib:juno:pmt}:
\begin{itemize}
\item[-] about 5000 dynode PMTs, model R12860, from Hamamatsu
         Photonics;
\item[-] and about 15000 Micro-Channel Plate Photomultipliers (MCP-PMT),
         produced by North Night Vision Technology.
\end{itemize}
The Hamamatsu R12860 PMT is based on a "Venetian-blind" dynode structure,
while the NNVT PMTs use one micro-channel plate.
They need different voltage dividers.
Figure~\ref{fig:BASE_schematics} shows the electrical scheme;
it can be seen that the high voltage supply to the anode is positive,
while the photocathode is on ground.
The signal output is doubled and the
maximum signal amplitude is limited to about 8~V to protect the
consecutive electronics from over-voltage.
One side of the board is soldered direclty to the PMT pins.
The HVU is mounted on the other side of the board.

%
%
The high voltage is generated by the HVU, a custom module that converts
a 24~V DC voltage to a high DC voltage (HV) using a cascade of
half-wave doublers (Cockroft-Walton multipliers).
Such a system does not need any HV cables or connectors.
The module is equipped with an embedded microcontroller. It monitors
all operations and provides a RS485 half-duplex interface to the GCU.

The properties of the HVU are:
\begin{itemize}
\item[-] range of HV: 1500~V - 3000~V in steps of 0.5~V.
\item[-] ripple: 10~mVptp
\item[-] HV long term stability: 0.05\%
\item[-] temperature coefficient: 100~ppm/$^\circ$C
\item[-] maximum output current: $300~\mu\mbox{A}$
\end{itemize}

\subsection{Preamplifier and Analog-Digital Unit}

\begin{figure}[htb]
        \begin{subfigure}{.40\textwidth}
                \centering
                \includegraphics[width=1\linewidth]{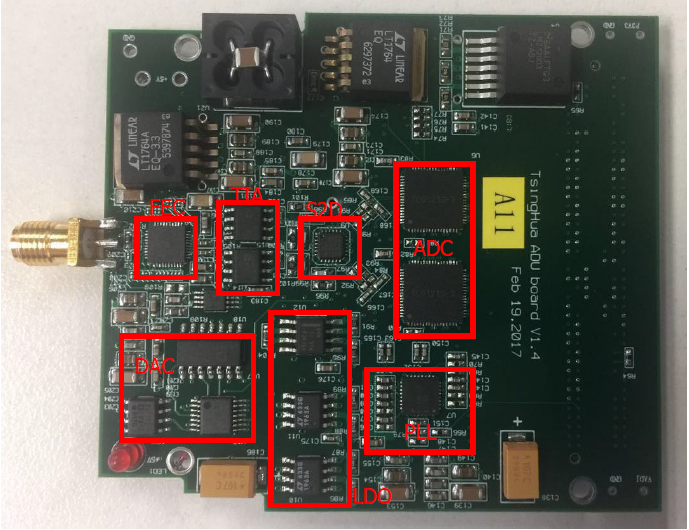}
                \caption{}
                \label{fig:ADU_front}
        \end{subfigure}
        \begin{subfigure}{.55\textwidth}
                \centering
                \includegraphics[width=1\linewidth]{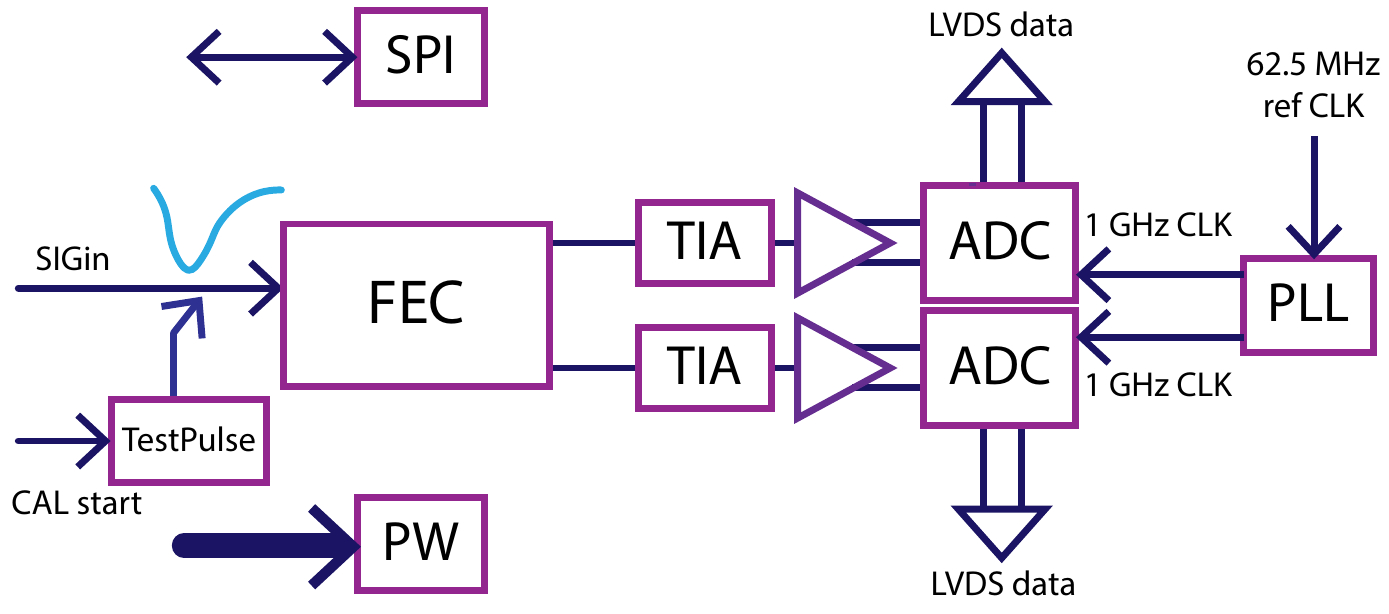}
                \caption{}
                \label{fig:ADU_scheme}
        \end{subfigure}
        \caption{\textit{(a): ADU top side; the main components have been
                 highlighted. (b): ADU logical scheme.}}
\end{figure}

To allow for maximum flexibility, during the design of the readout
electronics, it has been decided to mount an FMC~\cite{bib:fmc:standard}
low-pin-count connector of the GCU board\footnote{It can be seen on the
right part of \protect{Figure~\ref{fig:GCU_back}}.}, which allows to
'plug' different Analog-Digital Units during the prototyping and
testing phase.
The ADUs were mounted on an FMC mezzanine board
(see Figure~\ref{fig:ADU_front}).
The ADU receives the input charge, converts it into a voltage,
digitizes the waveform, and sends it to GCU for further processing.
The input signal is connected to the ADU thanks to an SMA connector.
As can be seen from Figure~\ref{fig:ADU_scheme}, the ADU consists of a custom
Front-End Chip (FEC), two commercial Trans-Impedance Amplifiers (TIA),
two drivers, and two custom ASIC ADCs. The circuit is completed by a
Phase Locked Loop (PLL) and some peripheral circuits.
Each ADC 
digitizes the analog signal at 1~Gsample/s
with a nominal resolution of 14 bits and an effective resolution of about
12 bit, taking into account the digitization noise.
The two preceding drivers amplify the signals with different gain:
a low gain with a dynamic range from 0 to
7.5~V  - equivalent to about 1000 pe - and a high gain with a reduced
range from 0 to 960~mV equivalent to 128 pe.
The output link uses a 14-bit Double Data Rate (DDR) parallel bus,
with the data synchronized to a 500 MHz clock.
This sampling clock is generated by an external Phase-Locked Loop
(PLL) mounted on the ADU. It receives the system clock of 62.5 MHz from the
GCU and provides a low jitter (100~fs RMS) 1~GHz clock to the ADC.
The circuit is completed by a Test Pulse Circuit, which can generate a
programmable test pulse to check the status of the full electronics chain.

\subsection{Global Control Unit}
\label{sec:BX:GCU}

\begin{figure}[htbp]
   \begin{subfigure}{.52\textwidth}
                \centering
                \includegraphics[width=1\linewidth]{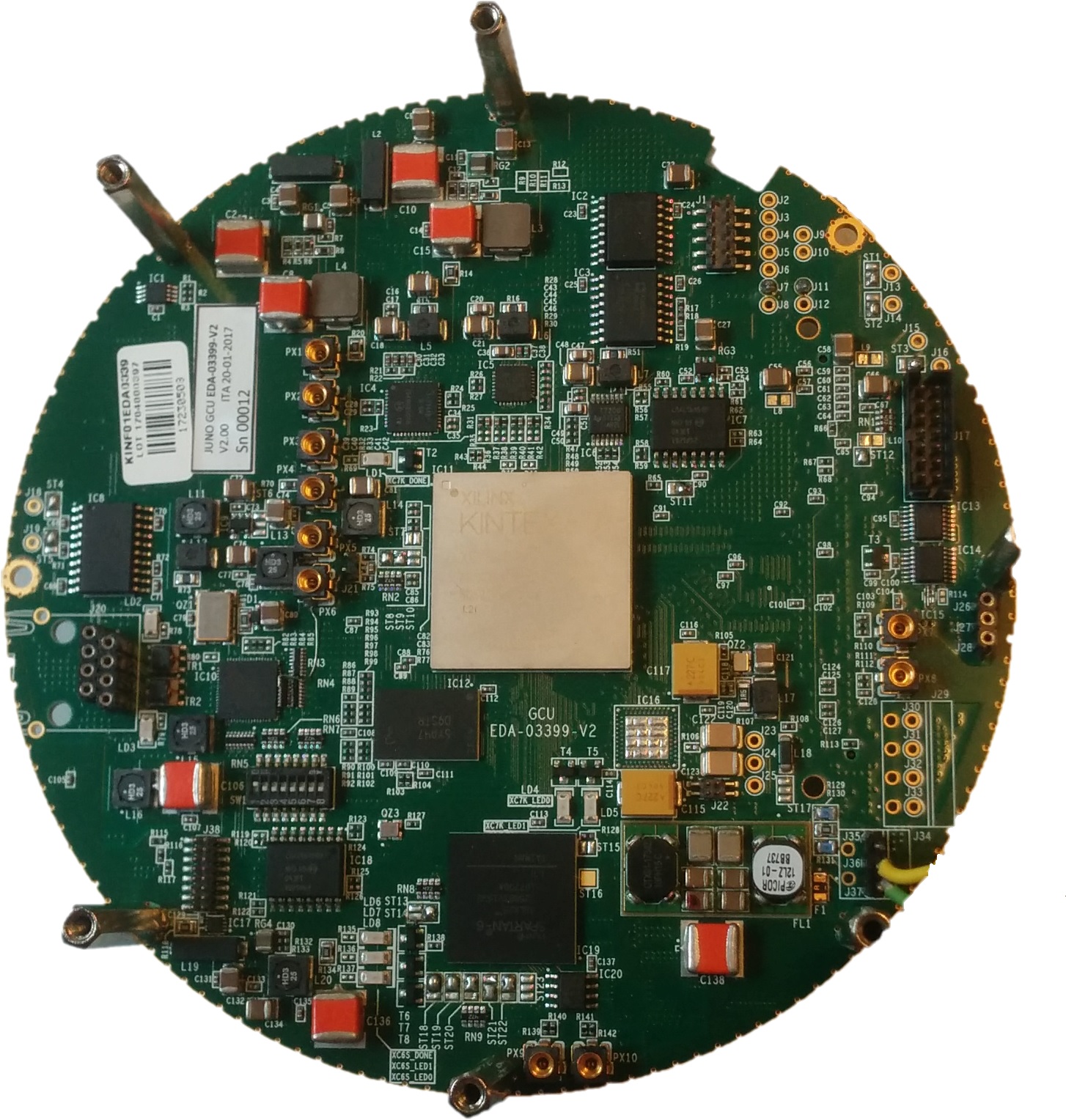}
                \caption{}
                \label{fig:GCU_front}
        \end{subfigure}
        \begin{subfigure}{.52\textwidth}
                \centering
                \includegraphics[width=1\linewidth]{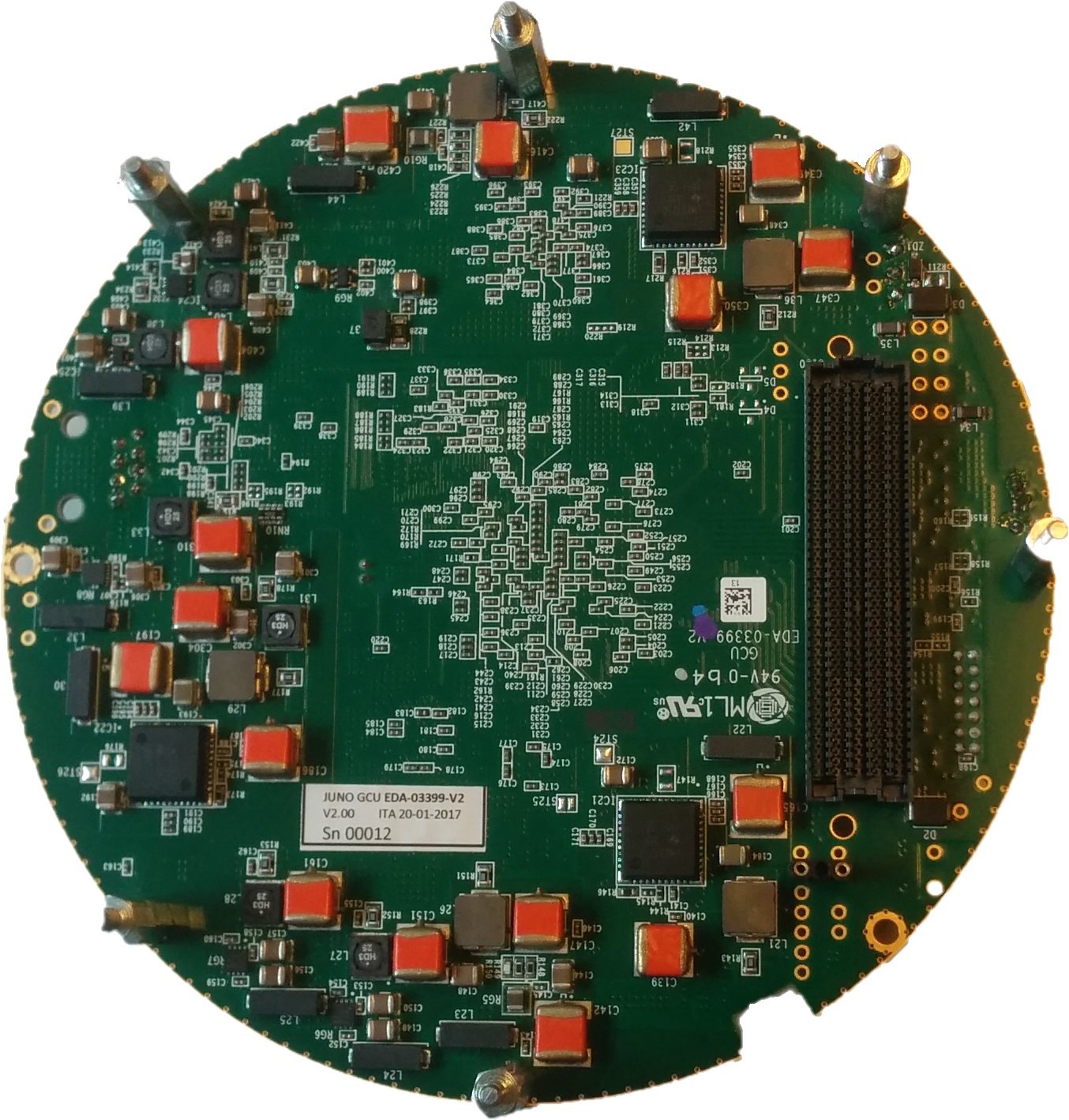}
                \caption{}
                \label{fig:GCU_back}
        \end{subfigure}
        \caption{\textit{GCU prototypes picture: top side (A)
                         and bottom side (B).}}
        \label{fig:GCU}
\end{figure}

The Global Control Unit (GCU) is the core of the JUNO readout electronics;
Figure~\ref{fig:GCU} shows a top (left) and bottom (right) photograph of
one of the GCU prototypes.
The main task is the acquisition of the PMT waveform, their processing
(local trigger generation, charge reconstruction, and timestamp
tagging) and temporary storage
before sending it to the data acquisition (DAQ) upon a  trigger request.

\begin{figure}
\centering
  \includegraphics[width=1\columnwidth]{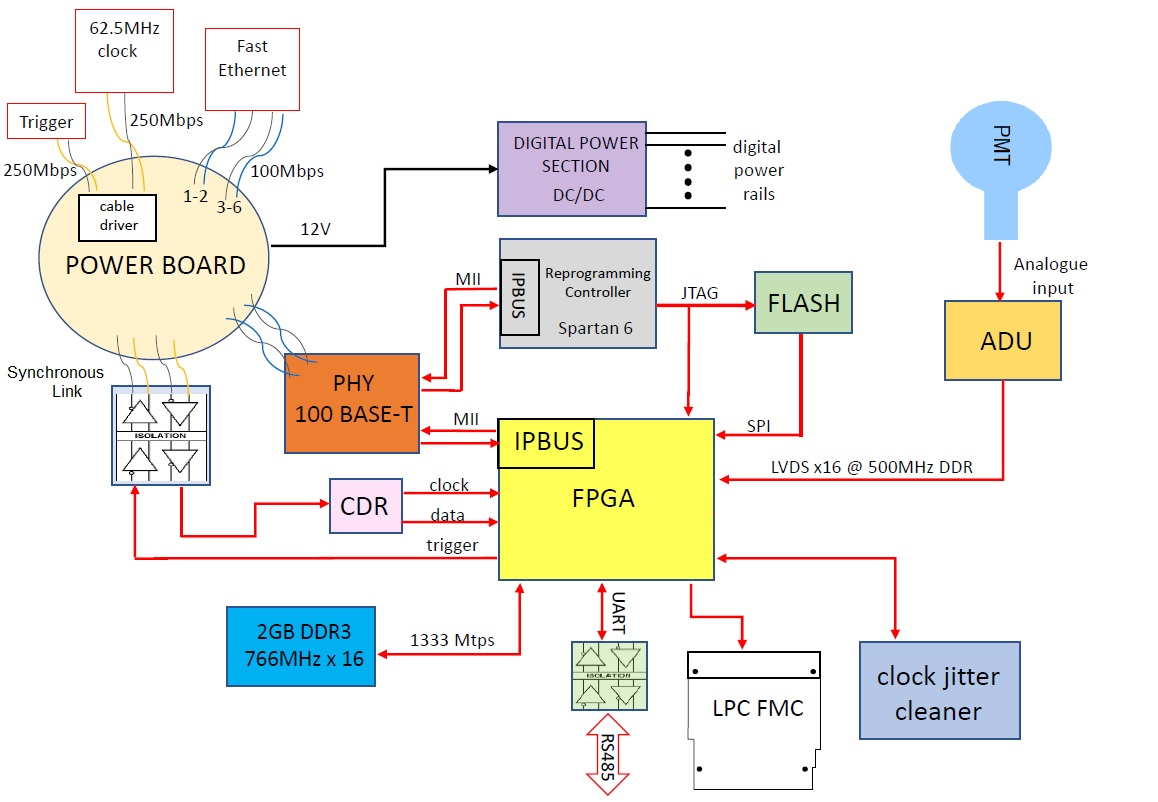}
  \caption{GCU Block diagram. A description of the different parts
           is given in the text.}
  \label{fig:blocks_GCU}
\end{figure}
A block diagram of the GCU can be seen in Figure~\ref{fig:blocks_GCU}.
The core of the board is a Xilinx Kintex-7 FPGA (XC7K160T), which is a good
compromise between number of available I/O ports, power consumption,
performance and cost.
A continuous stream of 14-bit data, sampled at 1 Gsample/s is
transferred from the ADC to the FPGA via 14 LVDS lines (500~MHz DDR)
The FPGA is able to
handle all the data packaging, processing and buffering. Metadata, containing
for instance the timestamp and the trigger number, is attached
to the event segments packages and stored in the board memory.
Upon a trigger request, validated waveforms are sent to the DAQ event
builder via Fast Ethernet.
The IPBus Core~\cite{bib:ipbus} protocol is used for data transfer,
slow control monitoring, and control operations.
It allows a transparent manipulation of the FPGA register across the Ethernet,
allowing to connect the Ethernet network to the I2C,
SPI, and UART GCU local buses.
A typical slow control operation is the setting of the PMT High Voltage
through the HVU, which is connected via an optically isolated RS485 interface
to the Kintex-7 FPGA; or the readout of the local GCU temperature sensors.
%
The synchronization and communication protocol, running on the two
synchronous links, is based on the Timing, Trigger and Control
(TTC)~\cite{bib:ttc} protocol, developed at CERN.
It provides the capability to exchange data between the GCU and the
BEC, such as trigger timestamps and calibration
information, as well as sending trigger input upstream to the Central
Trigger Processor (CTP).
A description of the TTC implementation on the current hardware and
discussion of the results can be found elsewhere~\cite{bib:ttc:gcu}.
The data streams are DC-balanced. A Clock Data Recovery (CDR) chip in the
GCU recovers the master clock of the experiment from the data stream.
The synchronisation is a key feature. It guarantees that all the 20000
local clocks are aligned with the global time within a system
 clock period of 16 ns.

A critical point in the readout scheme is the capability to handle
8~Gbit/s  of raw data, (14 LVDS lines at 500~MHz DDR),
from the ADC, continuously.
The waveforms need to be stored while waiting for triggers from the CTP.
We expect a trigger latency of about $100~{\mu}s$.
Upon a trigger, a readout window of pre-defined length will be extracted from
the local buffer and sent to DAQ through the asynchronous link.
A circular, level-1 cache is allocated inside the main
FPGA memory. The available block RAM in the Kintex-7
is 11700 Kbits and it allows to store up to $1.4~\mbox{ms}$ of data.
which is well above the required latency.

In normal operation mode a trigger rate of about 1~kHz is expectd.
In case of a supernova explosion, the data rate will rapidly increase by
orders of magnitude.
The FPGA's internal cache will be too small to handle the data.
Therefore, a dedicated 2~GByte DDR3 memory has been added to the GCU.
The memory controller supports write operations up to about 21.3~Gbit/s
which is sufficient to handle the incoming data rate and to store two
seconds of continuous data.
The usage of a data compression algorithm would further improve the
effectively available memory.
Since the GCU will no loger be accessible after the detector will be
filled with water and liquid scintillator, the only interface,
Fast Ethernet, has to provide both data readout and remote FPGA
reprogramming.
Therefore, the GCU is equipped with a second smaller FPGA (Spartan-6)
with the purpose of ensuring a fail-safe reconfiguration of the Kintex-7, by
means of a virtual JTAG connection, over the IPbus, eliminating the need
of a dedicated JTAG connector and cable.
As can be seen in Figure~\ref{fig:blocks_GCU}, the two FPGAs
are connected to the Physical 100 BASE-T Ethernet switch and interconnected
via JTAG. The virtual JTAG also allows to use the Xilinx
debugging tools (Impact and Chipscope). A custom
Xilinx virtual cable server, XVC~\cite{bib:xvc}, opens a TCP port for the Xilinx
tools and provides support for the IPbus/UDP protocol bridging the JTAG
commands to the GCU's JTAG chain via fast Ethernet, passing throughout the
IPbus core instantiated in the Spartan-6.

\subsection{Power and Communication Board}
\label{sec:BX:PB}
The Power and Communication Board (PB) provides the power to the 'wet'
electronics and the interface to the CAT-5e cable that connects to the
'dry' electronics.
Power is transmitted through the asynchronous data links using a custom
Power Over Ethernet (POE) approach:
the standard POE~\cite{bib:poe_standard} technology is adopted
for the power rails, but with a lower voltage (24~V instead of
48~V\footnote{Due to high reliability design constraint.}) and without
the overhead of the POE protocol.
Analog power is conveyed through the clock link. The CLK signal is AC
coupled onto a power rail.
The voltage of both power lines can be adjusted independently to compensate
for the power losses over the long 100~m cable.

The PB is connected to the GCU. Data links and
a dedicated 12~V (1~A, max) power rail are provided. From the 12~V power
rail, the GCU will generate, internally, all the required voltages.
The PB also connects to the HVU through a low ripple power line.
The voltage is in the range between 23~V and 30~V with a maximum allowed
current of 80~mA.
Three seperate ground potentials are provided. They are connected to the
'dry' electronics through the shields of the corresponding CAT5e cable.
There is a digital ground and an analog ground which are connected to each
other at a single point in the 'wet' electronics in the ADU.
A third ground is transported on the outer shield of the CAT5e cable and
connected to the steel housing of the 'wet' electronics for electrical
shielding. 

\begin{figure}[htb]
\centering
\includegraphics[scale=0.45]{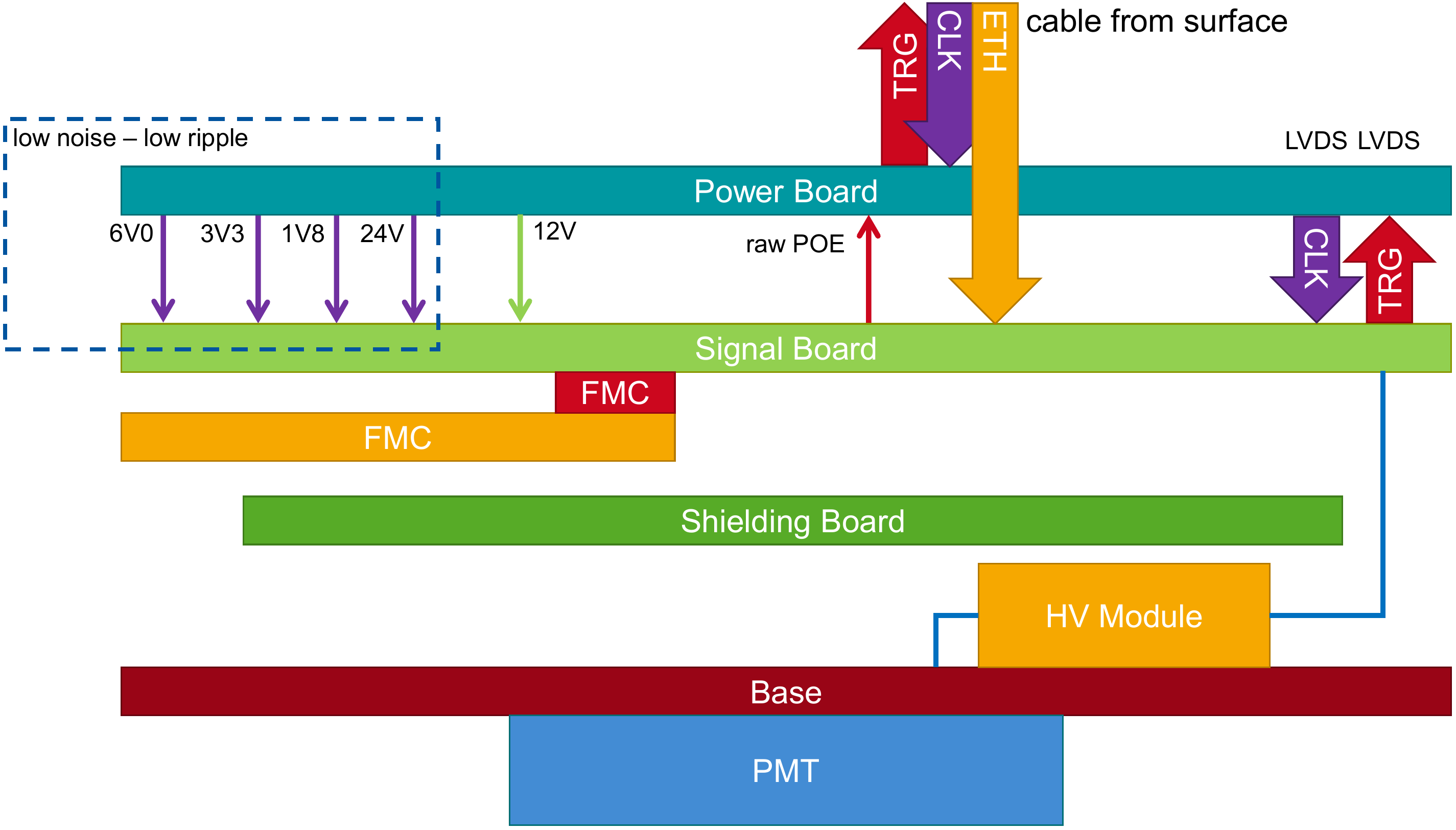}
\caption{Electronics boards connections schemes.
The different signals and power rails routing is indicated.
 Details are given in the text.}
\label{fig:BX:PB:connections}
\end{figure}

For the assembly of the boards, the CAT5e cable is soldered onto the
PB. Cable ties are foreseen to hold the cable in place to protect the
solder joints from possible stress. The cable will be split
into its pairs, which are then soldered close to the
corresponding driveris/receivers
located in different positions on the PCB.
A scheme of assembly of the three boards with
the signal and power connections is given in Figure~\ref{fig:BX:PB:connections}.

\subsection{Back End Card}
\label{sec:BX:BEC}

\begin{figure} [h!]
\begin{center}
  \includegraphics[width=0.65\linewidth]{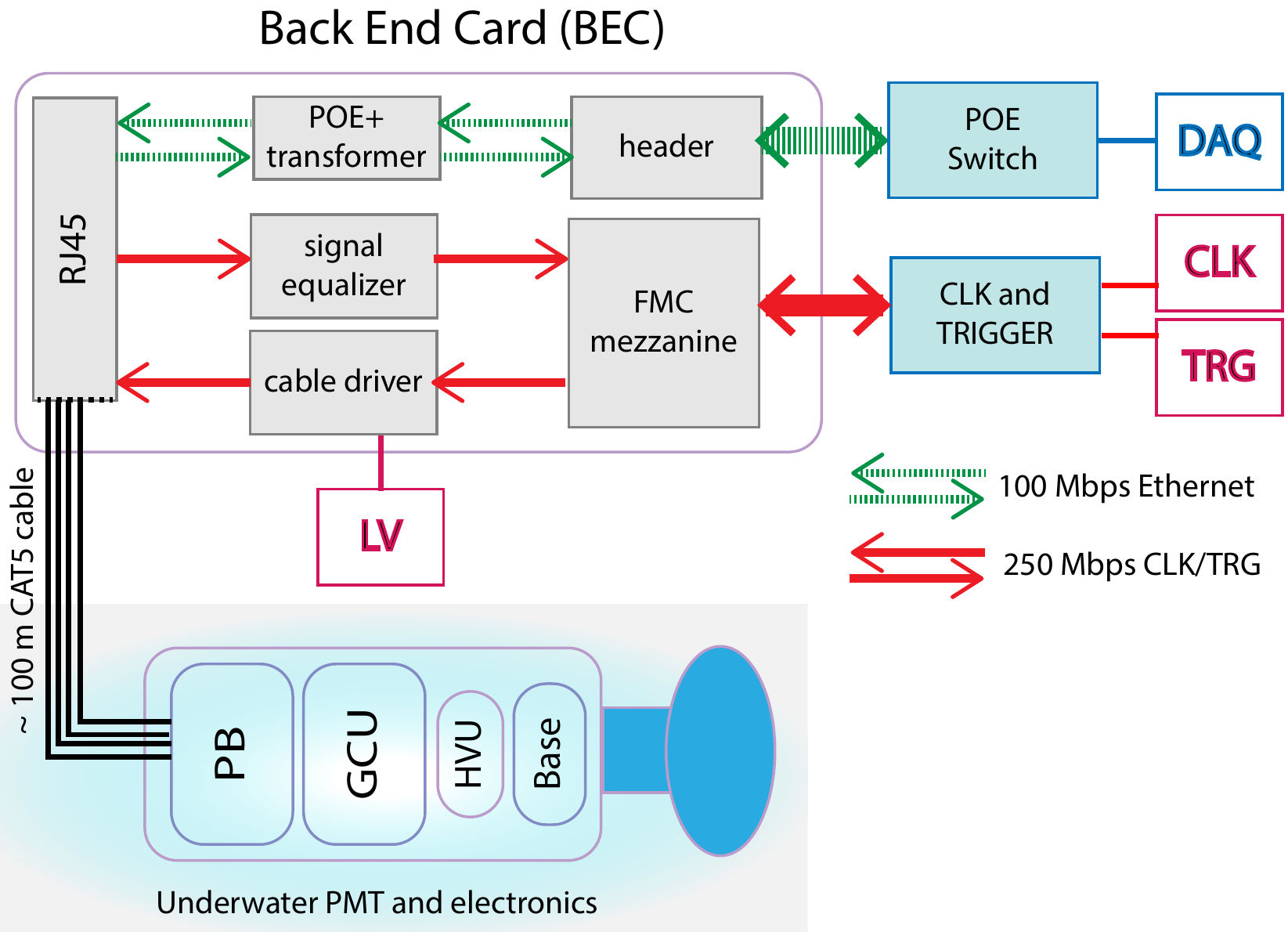}
  \caption{Logical diagram of the JUNO large PMT electronics, with
           BEC logical scheme enlightened.}
  \label{fig:bx_with_BEC}
  \end{center}
\end{figure}

The Back End Card (BEC) is the first board of the 'dry' electronics.
It is used as a concentrator and a bridge between the 'wet'
electronics and the DAQ and trigger systems.
The main task of the BECs is the handling of the data links
from/to the reception of the trigger input and the distribution of the
power and the clock to the 'wet' electronics.
One BEC connects to 48 GCUs.
Since JUNO will deploy around 20000 large PMTs, about 420 BEC will be needed.
A schematic view of with a focus on the role of the
BEC is presented in Figure~\ref{fig:bx_with_BEC}.

\begin{figure} 
\begin{center}
  \includegraphics[width=0.5\linewidth]{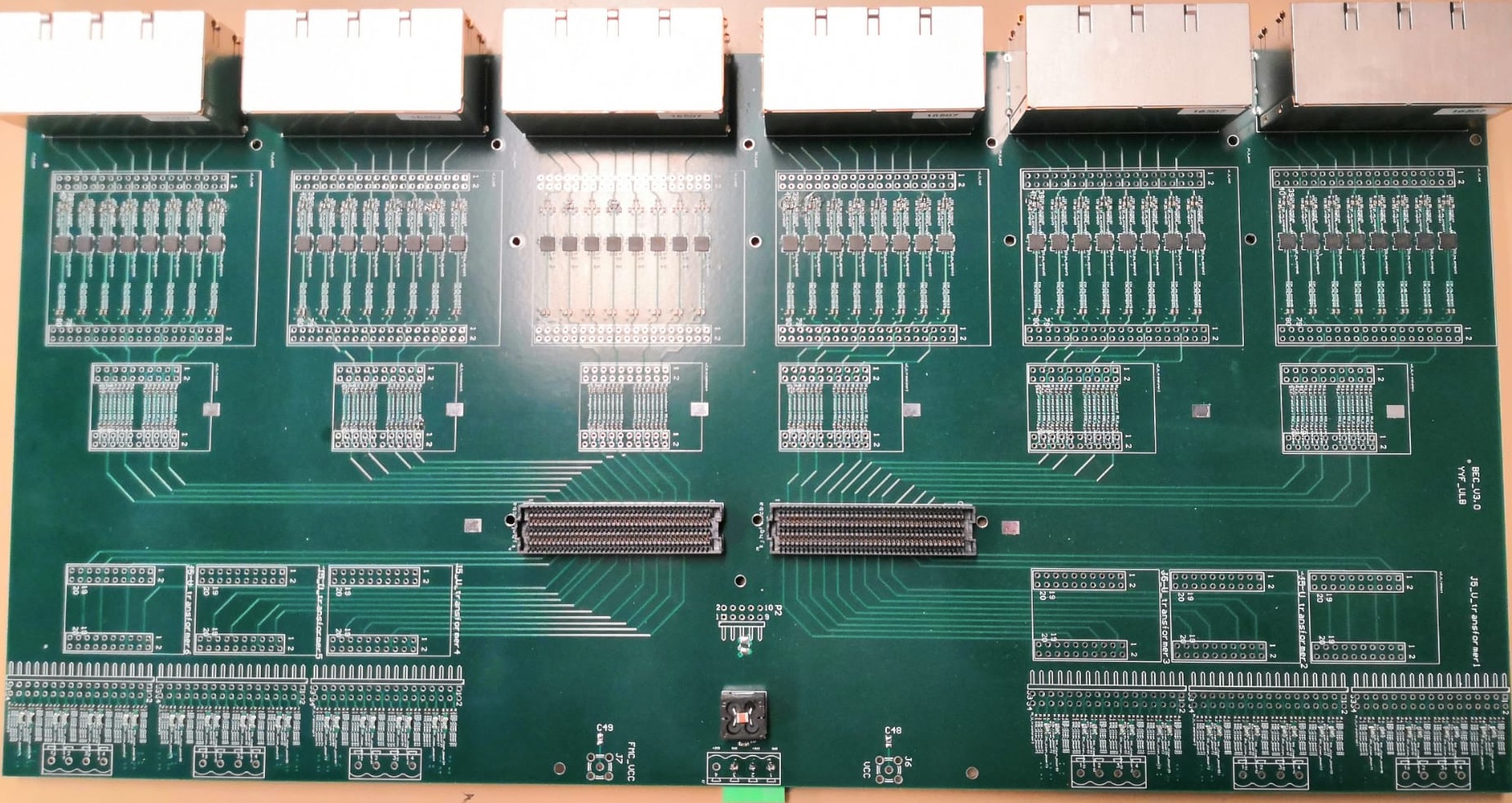}
  \includegraphics[width=0.45\linewidth]{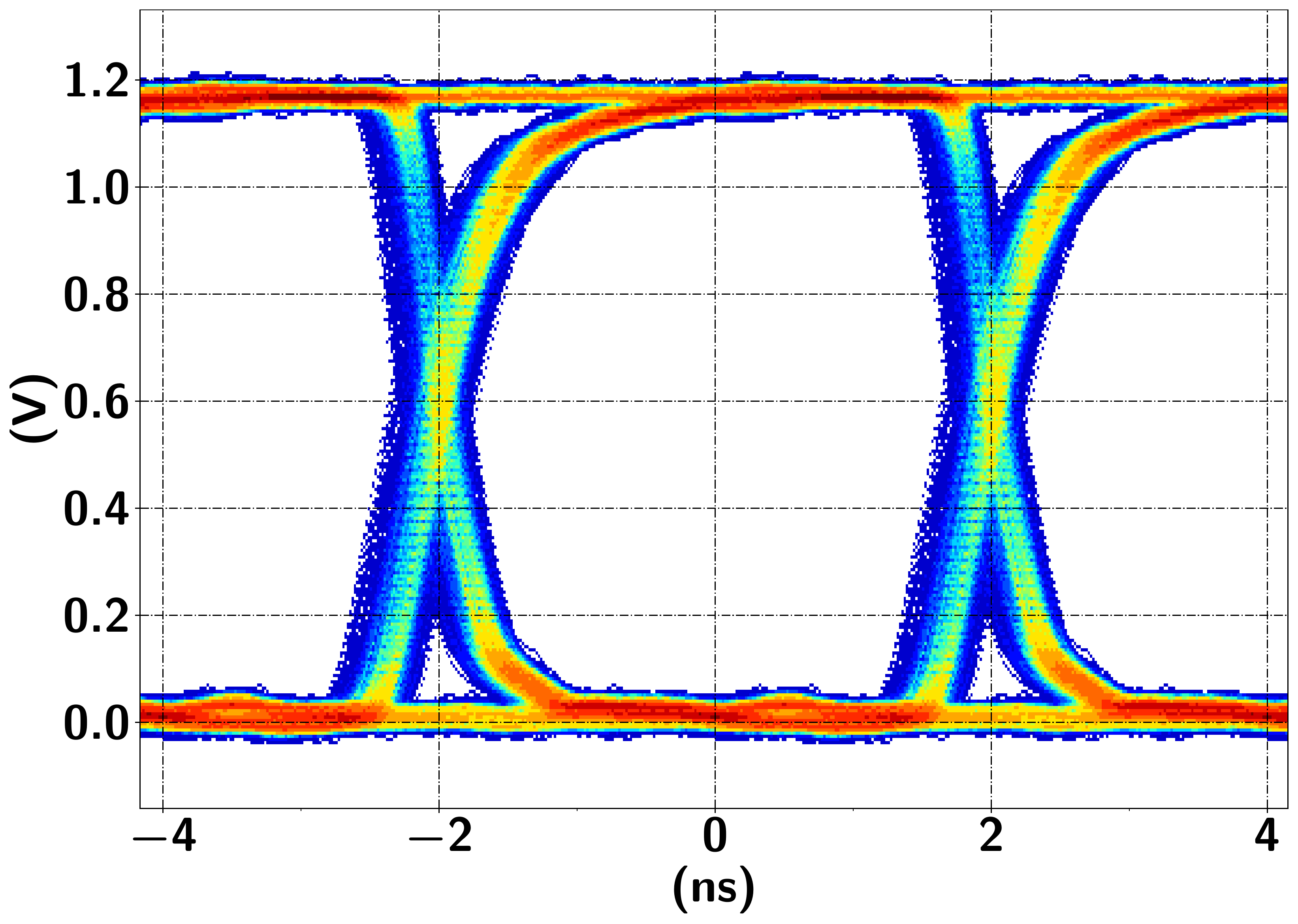}
  \caption{Left: BEC baseboard design. Right: Synchronization tests results,
           eye diagram.}
  \label{fig:BEC_design}
\end{center}
\end{figure}

The BEC consists of two parts: the baseboard and the
Trigger and Timing (TTIM) FMC mezzanine card.
The baseboard routes all the signals. It compensates the losses due
to the long cables on the incoming signals and connects to trigger,
DAQ system, central clock and power supplies.
The readout and slow control data streams which are transmitted over
Ethernet, are passively routed to a commercial POE switch.
The BEC baseboard design is shown in Figure~\ref{fig:BEC_design} (left part).

The PCB is equipped with 48 RJ45 connectors located
on the bottom side of the baseboard to provide the connections to the
'wet' electronics.
Close to the connectors, 48 equalizers are mounted to handle the
upcoming trigger inputs.
%
The output form the 48 differential pairs is connected to two
custom-defined LPC connectors with two serial 0 Ohm resistors in each path.
The two LPC connectors are situated in the middle part of the baseboard,
and provide connection to the TTIM. On the top side of the LPC,
another 48 differential pairs connect back to the RJ45 connectors
for the down-link trigger validation signals.
In total 96 differential pairs are connected to the two LPC connectors.
In the middle of the top part of the baseboard,
there is the power connector for the BEC itself.
It is separated from the power supplies for the 'wet' electronics
to allow for  flexibility in the grounding.
Since one BEC has 48 identical ports and each port supports
bi-directional data transfer,
two ports can be cross-connected for testing.
The TTIM can be used to generate 250 Mb/s PBRS data. The eye
diagram shown in Figure~\ref{fig:BEC_design} (right part), shows a
stable bi-directional data transfer realized connecting two channels
on a BEC board through a 100~m long Ethernet cable.

\section{Reliability}
\label{sec:reliability}
\begin{figure}[ht]
 \begin{center}
  \includegraphics[width=0.85\textwidth]{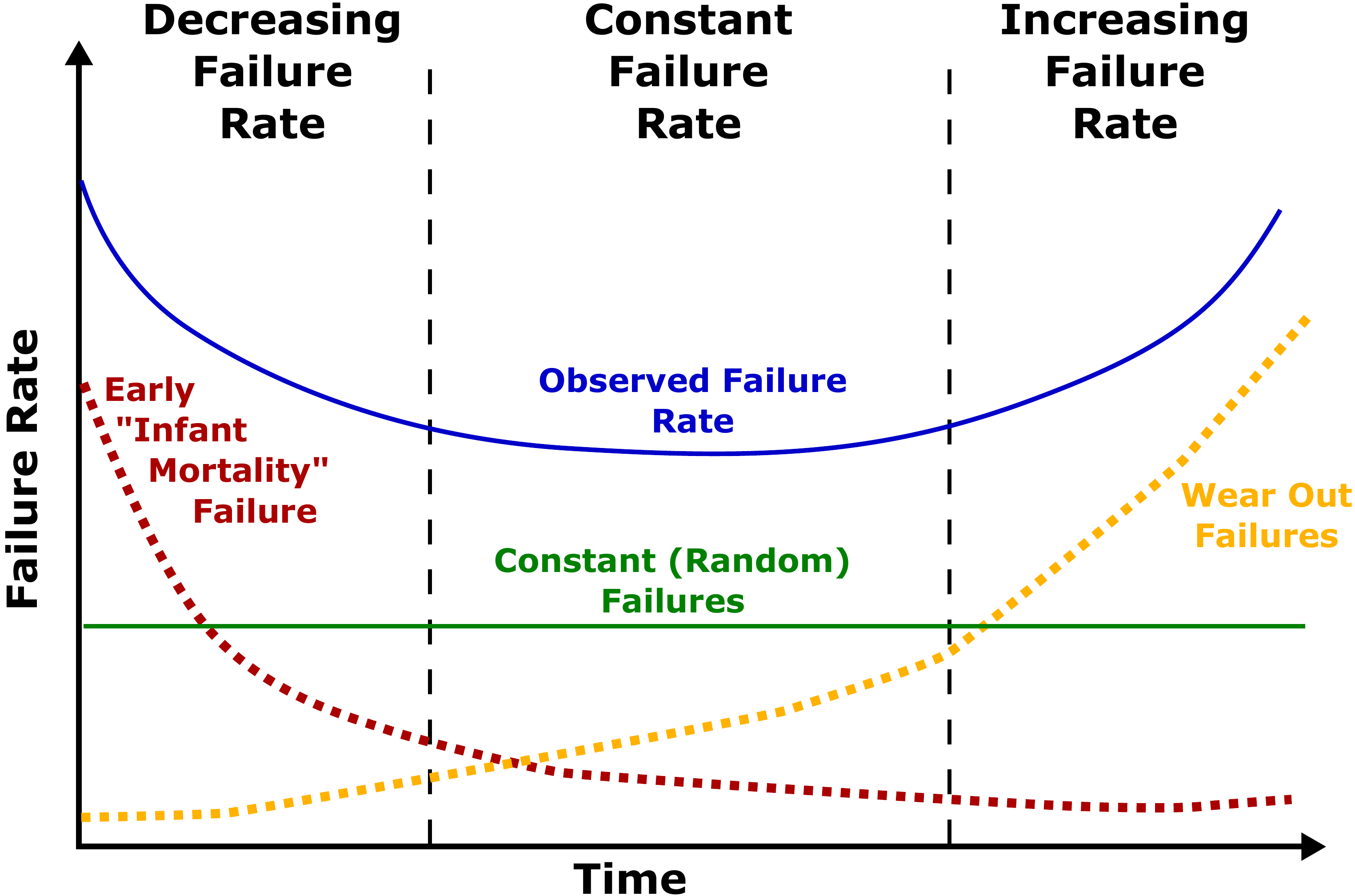}
  \caption{Development of the failure rate throughout the lifetime of
           an electronic component~\cite{bib:reliawiki}.
 \label{fig:bathtube_curve}}
 \end{center}
\end{figure}

The 'wet' electronics cannot be accessed after liquid scintillator filling.
As mentioned in the introduction, JUNO requires less than 1\% of the
channels should fail
during the first six years of operation. We assume that half of the failures
stem from PMTs and their bases, so that less than 0.5\% of the electronics
may fail, taking also into account failures of the cables and leackage
into the electronics housings.
%
The failure of electronics over time can be described by three
major phases in the so-called bathtub curve
(see figure \ref{fig:bathtube_curve}).
In the beginning of the operation, the failure rate is dominated
by infant mortality.  During this phase, devices or components with
small defects, like bad solder joints, fail.
For high reliability electronics infant mortality can be overcome with
carefull screening and burn-in. Throughout the useful live-time of a device
random failures are dominant, leading to a constant failure rate.
All discussions and definitions in the following sections describe
this random dominated lifetime.
At the end of the lifespan the risk increases again
due to aging effects like decreasing chemical stability \cite{bib:reliawiki}.
In table \ref{tab:acronyms} the relevant acronyms used in reliability
engineering are specified.
The essential value is the failure rate $\lambda$,
expressed in failures in time (FIT).
It is assumed to be constant over the useful lifetime.
The probability to fail can be calculated using an exponential function
(eq.\ref{eq:failrate}):

\begin{equation}\label{eq:failrate}
        P(\text{fail}) = 1-e^{-\lambda\cdot t}
\end{equation}

The failure rate $\lambda$ is usually normalized to $10^9~\mbox{hours}$
of operation, which shifts typical electronics to
FIT-values of $\mathcal{O}(1)$.

\begin{table}[ht]
\caption{Definition of acronyms used in reliability
         engineering~\cite{bib:reliawiki}.
\label{tab:acronyms}}
\begin{tabular}{|l|p{7.5cm}|}
\hline
Terms & Definition \\
\hline
Failure Rate $\lambda$ & The failure rate describes number of failures per
time for one component, assuming a constant failure rate.
$\lambda$ is given in units of FIT. \\
\hline
Failure In Time (FIT) & Measure of the number of fails per device 
$10^{9}$~hours,
e.g. $\lambda = 100~\mbox{FIT} = 100$ failures in $10^9\mbox{h}$. \\
\hline
Mean Time To Failure (MTTF) & The Mean Time To Failure is the mean
lifetime under operation before a defect occurs and is consequently the
inverse of the failure rate $\lambda = \frac{1}{\text{MTTF}}$.
Mean Time between Failures (MTBF) is a synonym if the device is repairable.\\
\hline
\end{tabular}
\end{table}

\subsection{Calculating the reliability}
A device's failure rate can be described by the sum of the failures of
all included components.
The military handbook MIL-HDBK-217F~\cite{bib:MIL-HDBK-217F} Notice 2
was used as our baseline and the FIDES~\cite{bib:FIDES}
served as a cross-check.
The military handbook is a well established tool for estimating the reliability
of a device.  It is based on data obtained during operation and uses simple
assumptions to create easily usable models.
For the reliability calculation, two different methods are introduced for
different stages of the project: the "part count" and the "part stress" method.
The part count method is a conservative approach that can be used in the early
phase of a project to get an initial estimate of the reliability.
The part stress method refines this estimate at a later state of the
development, when all part parameters, e.g. voltage stress and temperature,
are known.
The results are conservative but reasonable for most devices and
components~\cite{bib:Fides_exp}.
However, for some components with significant improvements in processing
over the last few years, like CMOS-microcircuits, the reliability results
are too negative.
Additionally, SMD components are missing in these models, but they play a
crucial part in modern electronics.
On top of the failure rate of the components, we need to consider failures
of the PCB assembly. It is calculated with the FIDES guide~\cite{bib:FIDES}.
The failure rate depends mainly on the technology, the number of solder
joints, the environment of the final assembly, and the reliability
of the manufacturer.

One may either test every component individually or the entire device
in a single measurement. But with the entire device, the problem arises
that a failing component may lead to a cascade of other components
failing and the origin of the failure may not be identified.
Alternatively, testing all components by themselves is a valid
method too, but as the failure rate of standard components is very low,
many components and a long testing time are needed.
A common way to accelerate the tests is to increase the stress on the
component, for example increasing the temperature to accelerate chemical aging.
The simplest way to describe the probability of a device to fail is
by an exponential function. Some assumptions have to be made.
The failure rate of the device has to be constant,
which is valid only after infant mortality and before being worn out.
The failure rate can be calculated as
\begin{equation}
        \lambda = \frac{\chi^2(2\cdot(f+1), \text{CL})\cdot 10^9\mbox{h}}{2\cdot t\cdot d\cdot AF}
        \label{eq:lambda2}
\end{equation}
Where, $\lambda$ is the failure rate in $10^{9}~\mbox{hours}$,
$f$ is the number of devices which failed, $\chi^2$ is the $\chi^2$ value for
$(2\cdot (f+1))$ degrees of freedom, given a confidence level CL. Finally,
$t$ is the test duration in hours, $d$ is the total number of devices
tested, and $AF$ is the acceleration factor, defined for thermal stress, as
in eq.(\ref{eq:lambda2}):
\begin{equation}
 AF = \exp\left(\frac{E_a}{k_B}\left(\frac{1}{T_{\text{use}}}-\frac{1}{T_{\text{stress}}}\right)\right)
\label{eq:lambda3}
\end{equation}
Here the activation energy $E_a$, in eV, the Boltzmann's constant $k_B$ and
$T_{use}$ and $T_{stress}$ are the absolute temperatures (in Kelvin)
of the accelerated test and normal use, respectively.
If a large number of failures is observed, the
$\chi^2$-function may be approximated by the number of failures,
but, usually, the number of failures is small.
Typically, a confidence level of $60\%$ is used.
The factor of $10^9~\mbox{h}$ normalizes the result.
During the test all device have to be operational, i.e. under power.
The early failures result from the defects that occur in production and
assembly. They need to be subtracted from the calculation.
We forsee a screening for early failures with some thermal cycling
to suppress infant mortality.
The target value for all of the 'wet' electronics is 95~FIT.

\subsection{Power and Communication Board reliability}
\begin{figure}[ht]
  \centering
  \includegraphics[width=0.45\textwidth]{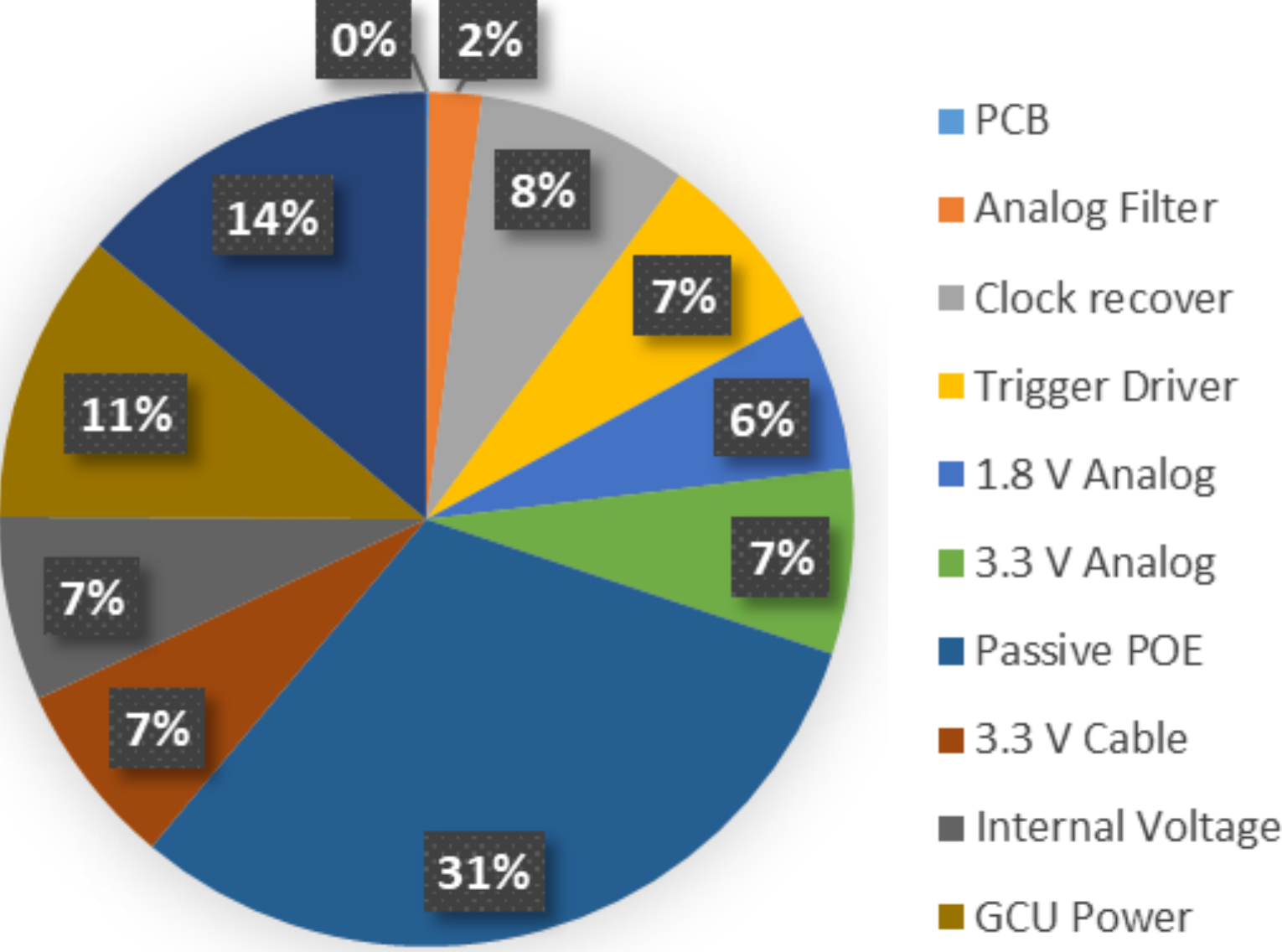}
  \hfill
  \includegraphics[width=0.45\textwidth]{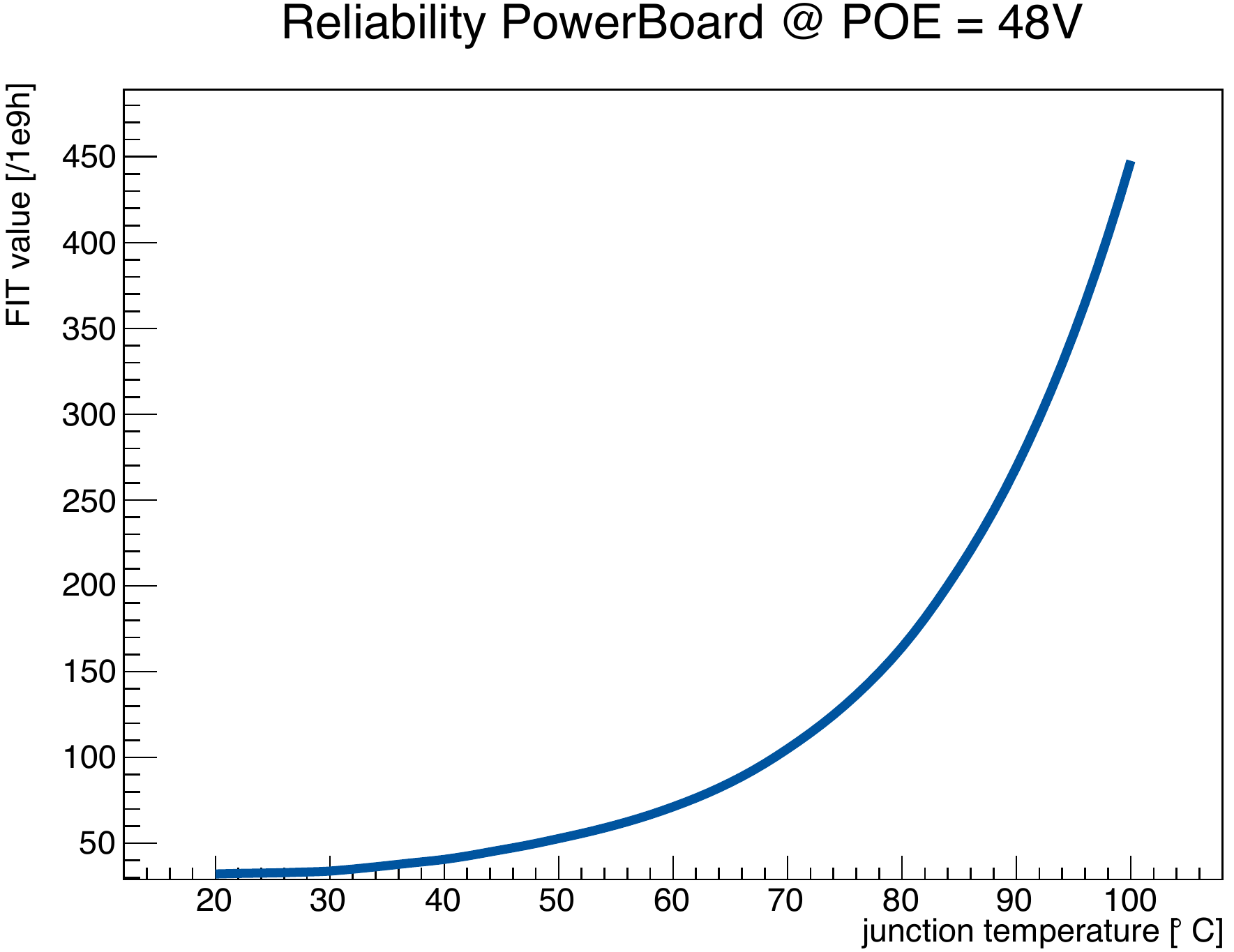}
  \caption{Power and Communication Board reliability. Left: single components
           contribution. Right: FIT value temperature dependence.}
  \label{fig:reliability_pie}
\end{figure}

As an example, the details of the reliability calculation for the PB
are presented below.
We modified the design and especially the selection of the components
through several interactions to minimize its failure rate.
It was decided to use only components which are qualified by the manufacturers
and FIT values are provided.
We use a conservative approach.
All components are classified as critical for the operation of the board.
The failure of a temperature sensor is assumed to have the same impact as the
failure of a truly critial component like the Ethernet transformer.
A dedicated code, \textit{ReliabilityCalc}\footnote{Available from RWTH Aachen
University through https://github.com/JochiSt/ReliabilityCalc,
DOI 10.5281/zenodo.1134161}, was developed.
The program calculates the reliability using the manufacturer's data or the
military handbook, including temperature dependencies and stress levels.
The failure rate of the PB with all of its 266 components is estimated at
\begin{equation*}
        \lambda < 40.4~\mbox{FIT}
\end{equation*}
at a temperature of $40^\circ\mbox{C}$ for every component.
The temperature was measured with a dummy board potted in oil,
with simple resistors simulating the heat dissipation.
The contribution of the different parts,
after optimization, can be seen in figure~\ref{fig:reliability_pie}.
The failure rate is dominated by one passive component:
the PoE coil~\footnote{Ethernet magnetics, 749012013, from W\"{u}rth
Electronics}.
Unfortunately, no alternative with a better failure rate could be found.
The right plot of Figure~\ref{fig:reliability_pie}
shows the FIT value as a function of
temperature. The exponential rise is dominated by silicon chips,
due to their high activation energy.
%

The failure of PB corresponds to about $42\%$ of the
budget available to all of the 'wet' electronics.
Considering the PB holds most of the power electronics this might be acceptable.

Table~\ref{tab:failures} presents the estimated failure rates of all of the
electronics. The major contribution given by capacitors is somehow
expcted due to a relative large number of capacitors on the board
(see Figure~\ref{fig:GCU}); moreover, many of those are tantalum capacitors
and they can't be replaced with ceramic capacitors due to their
unaffordable larger dimensions.

Obviously some more optimization is needed to fully reach the
goal of 95 FIT.
In parallel to the estimate of the failures in normal operation,
we investigated a number of exceptional events, such as power cuts.
We ensured that none of those exceptional events constitutes a significant
risk of failure.

\begin{table}[ht]
\begin{center}
\begin{tabular}{|c|c|l|} \hline 
Unit  & failure rate & Comment\\ \hline
HVU   &          50  &  dominated by the HV filter capacitor \\
GCU   &         107  &  dominated by  many capacitors\\
PB    &          40  &  see text\\
dry electronics &  0 &  replaceable\\
Cables &         30  & \\ \hline
\end{tabular}
\caption{Electronics estimated failure rates.
\label{tab:failures}}
\end{center}
\end{table}

%
%
%

\section{Prototyping and tests}
\label{sec:proto}
Several prototypes of all components of the 'wet' and 'dry' electronics
were produced. After extensive standalone tests the 'wet' electronics
was assembled into the stack seen in Figure~\ref{fig:BX:PB:connections}
and connected through a 100~m CAT5e cable to the prototype of a BEC.
Commercial units provided LV power and the clock signal to the BEC.
A preliminary version of the DAQ was used to communicate with the electronics.
The JUNO central trigger was not yet included in the tests.
Due to a mistake in the routing a cable was needed to patch the Ethernet
connection between PB and GCU. The sockets are visible in
Figure~\ref{fig:proto:bx_castle}. The 'wet' electronics was mounted on a
JUNO PMT (Hamamatsu R12860) to create a complete vertical slice of one channel.
The PMT assembly was located in a light-tight box. The vertical slice was
intensively tested, then the 'wet' electronics was potted into its watertight
housing and everything was tested again.
The results of the tests are presented below.

\subsection{Initial tests}
\label{sec:proto:test_bench}

\subsubsection{Linearity}
\label{sec:proto:ADC_linearity}
\begin{figure}[htb]
\begin{center}
\includegraphics[scale=0.42]{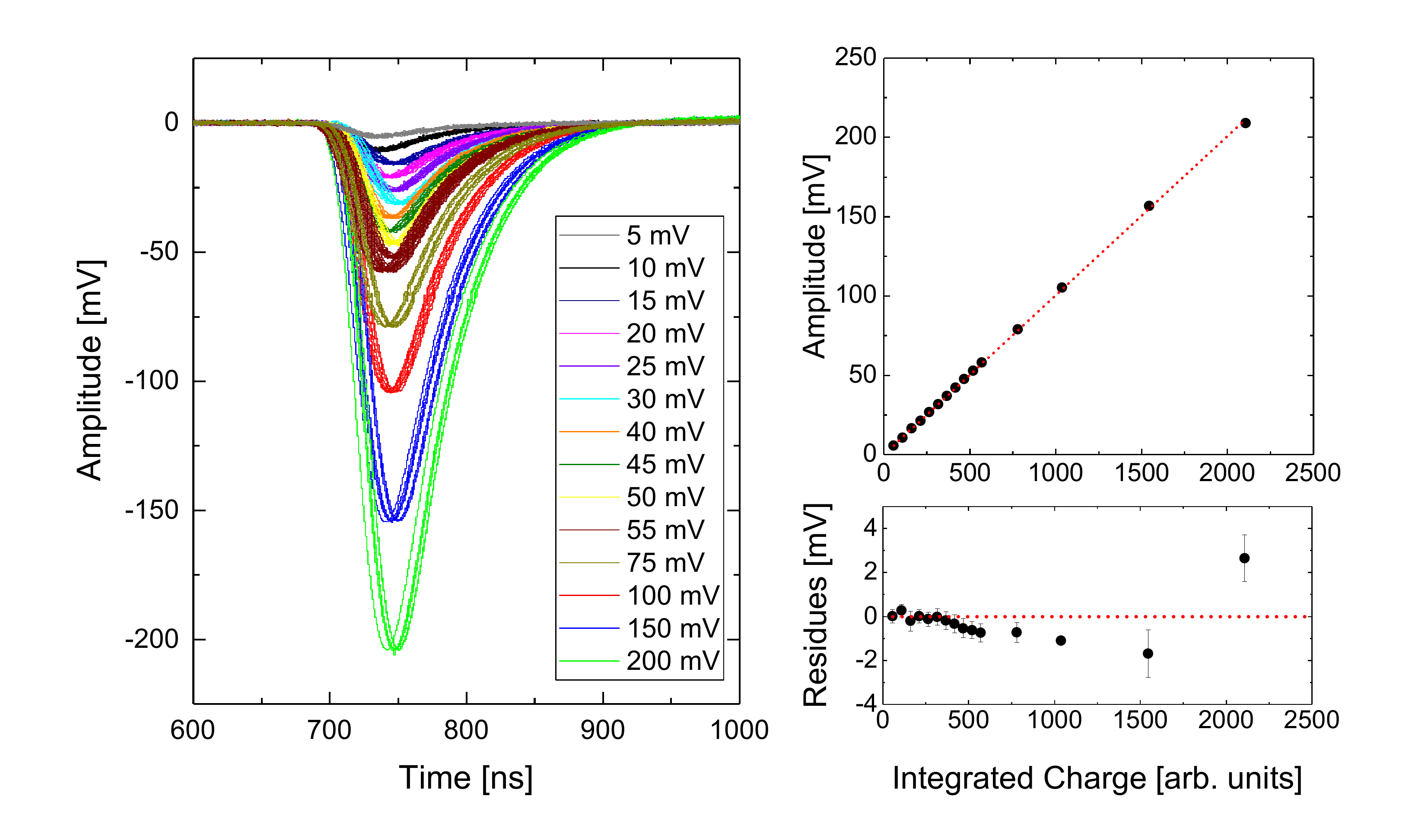}
\caption{Left: DT5810 input signals (smallest amplitude: 5~mV, highest
amplitude: 200~mV). Right: reconstructed charge as a function of the
input amplitude.}
\label{fig:proto:detector_emulator}
\end{center}
\end{figure}

To test the linearity of the response, the input was connected to
a CAEN Fast Digital Detector Emulator DT5810. It provides pulses
with a fixed rise and decay time, but with a programmable amplitude.
The left panel of Figure~\ref{fig:proto:detector_emulator} shows a sample of
simulated signals fed to the electronics: the amplitudes have been varied
from 5~mV up to 200~mV with a default rise and fall time of 30~ns and
120~ns, respectively.\footnote{The measurements were not performed for larger
amplitudes since the signal rise times increase dramatically going to
the $\mu$s domain.}
For reference, a single photon from the PMT creates a pulse with a typical
amplitude of 10 mV.
An external trigger, provided by the DT5810,
was used and data were acquired through the whole electronics chain.
The average charges of more than 10000 pulses per injected charge
are plotted on the right side of
Figure~\ref{fig:proto:detector_emulator}, against the input amplitude.
The plot shows excellent linearity.
The maximal deviation from a linear fit is 1.3 \%. The result is well inside the
JUNO requirements.

\subsubsection{Single photo-electron measurements}
\label{sec:proto:single_pe}
\begin{figure}[htb]
\centering
\includegraphics[scale=0.275]{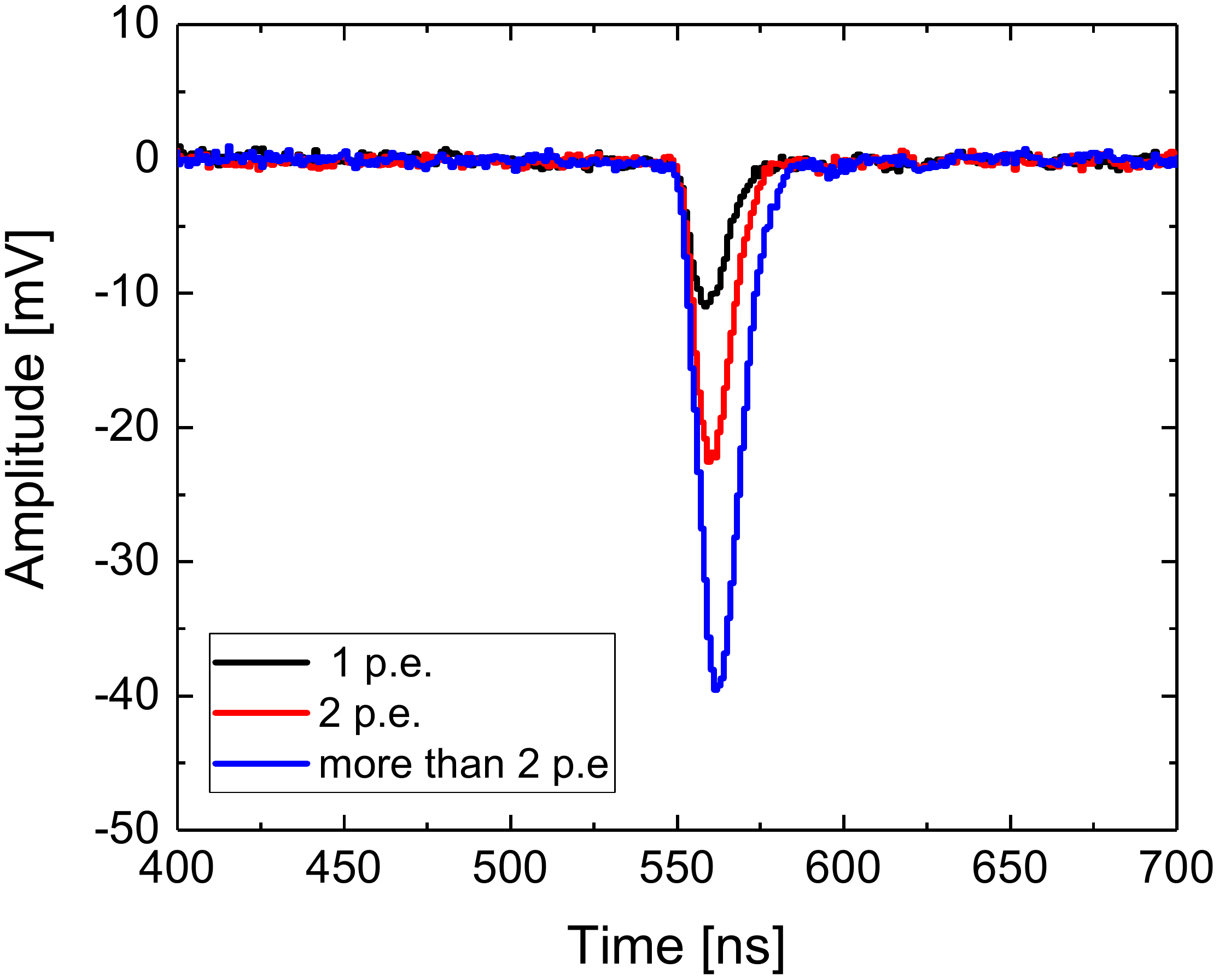} \hfill
\includegraphics[scale=0.4]{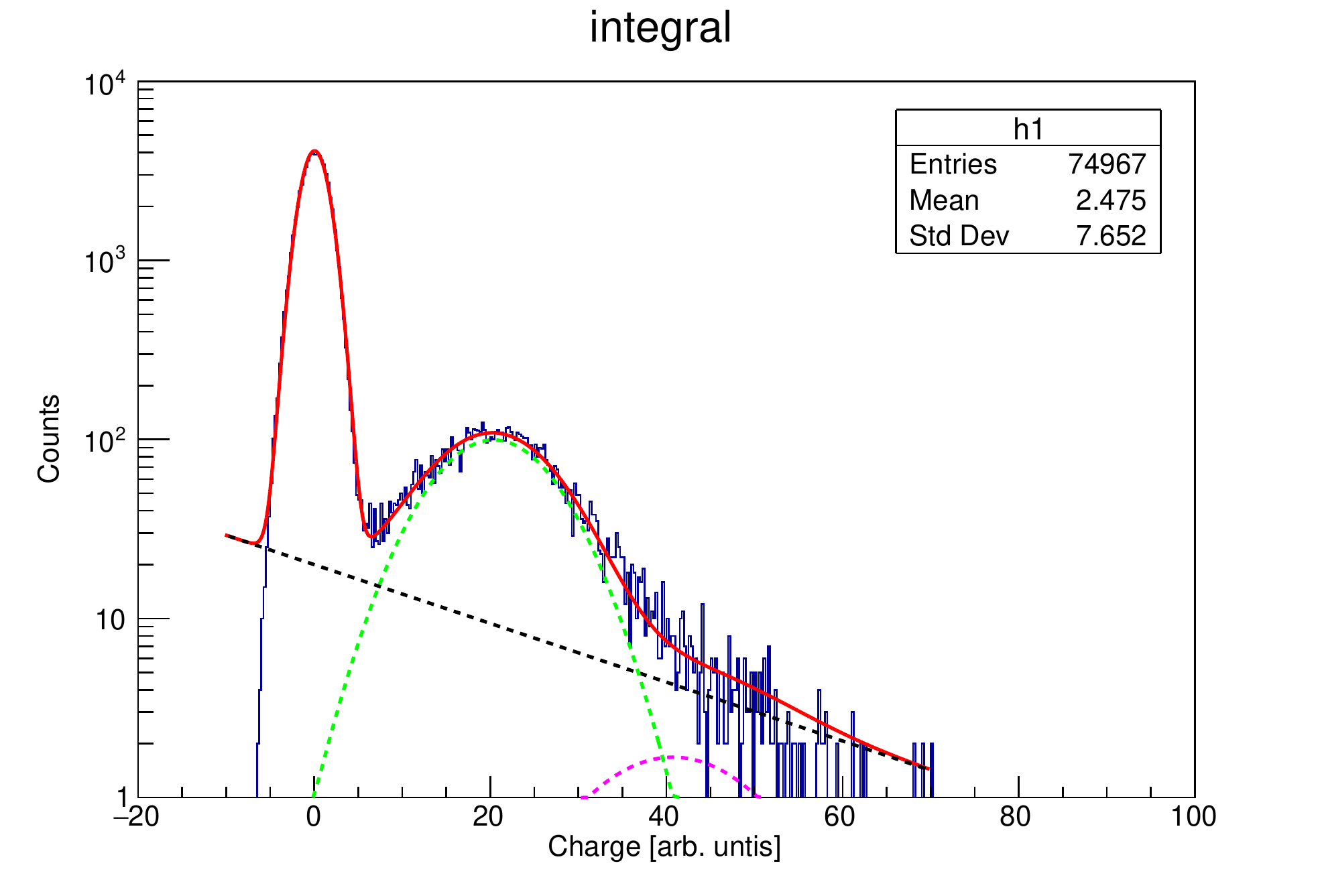}
\caption{Left: reconstructed pulses for different p.e. values.
Right: Single p.e. spectrum.
The spectrum was fitted with 3 Gaussians and an
exponential function for the background (black line): one Gaussian
for the noise peak centered at zero, one for the single p.e. contribution
 (green line) and the last one for the two p.e component (magenta line).}
\label{fig:BX:single_pe:PtV}
\end{figure}

Eventually the PMT HV was adjusted to a gain of $(1.75\pm 0.12)\cdot 10^{7}$.
The left plot of Figure~\ref{fig:BX:single_pe:PtV} shows a few pulses
of different amplitude recorded through the full vertical slice.
The data show a stable baseline with no overshoot
or wiggles at the tail of the pulses.
The rise time of the signal is around 7 ns and the decay time around 30 ns.
The pulses were integrated over 50 ns. The charge spectrum is shown in
Figure~\ref{fig:BX:single_pe:PtV}.
The mean amplitude for single p.e. was measured to be $9.39 \pm 0.03~\mbox{mV}$,
while the average noise level is $0.45\pm 0.04~\mbox{mV}$.
A signal-to-noise ratio of $20.9\pm 1.9$ was extracted for single p.e.
The fit presented in Figure~\ref{fig:BX:single_pe:PtV} gives
a peak-to-vally ratio of 3.8.
The single p.e. resolution is around $31 \%$.
The vertical slice was in stable operation for few days whithout any
loss of data.

\subsection{Tests after potting}
\label{sec:proto:test_final}
\begin{figure}[htb]
\centering
\includegraphics[scale=0.24]{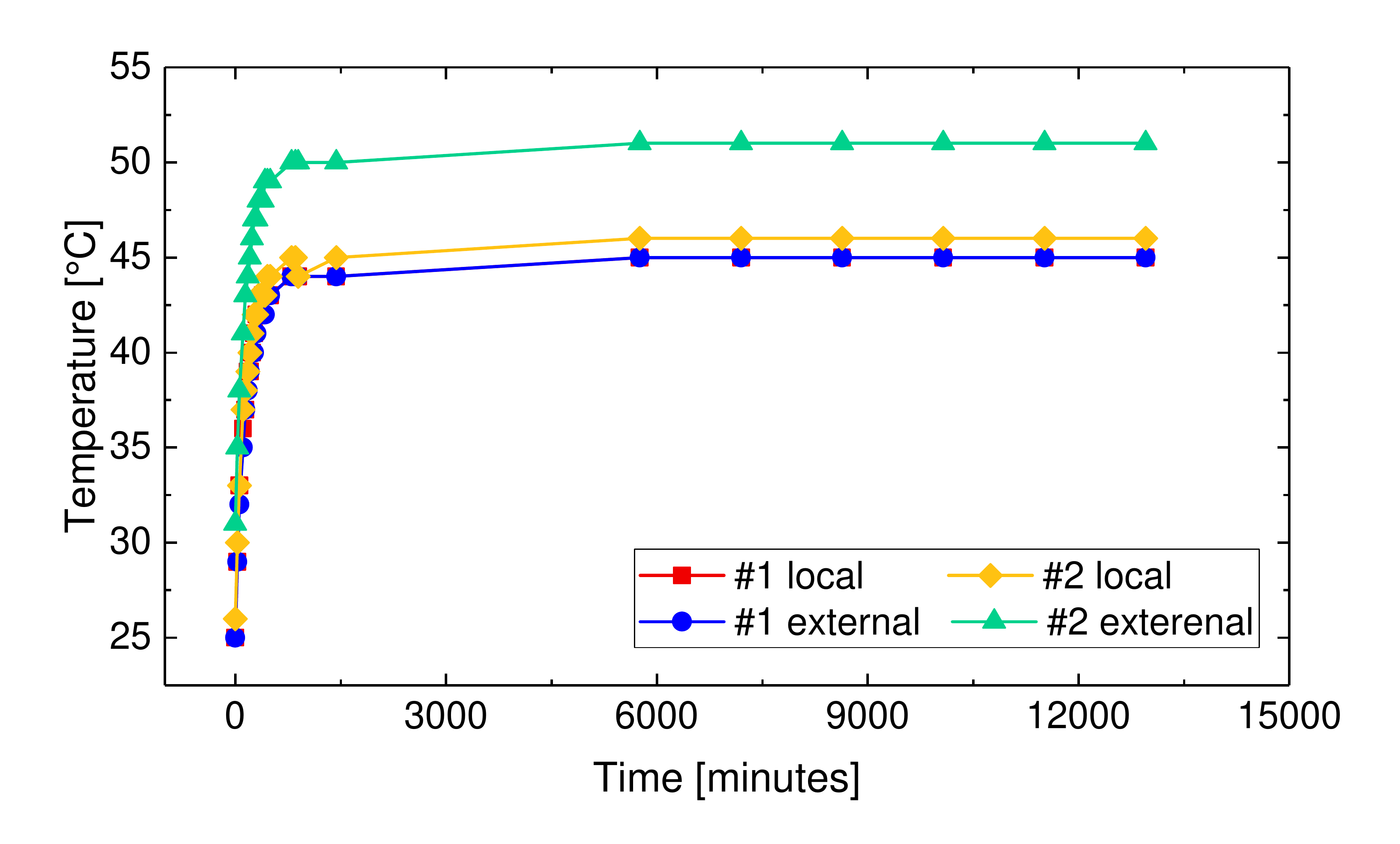} \hfill
\includegraphics[scale=0.25]{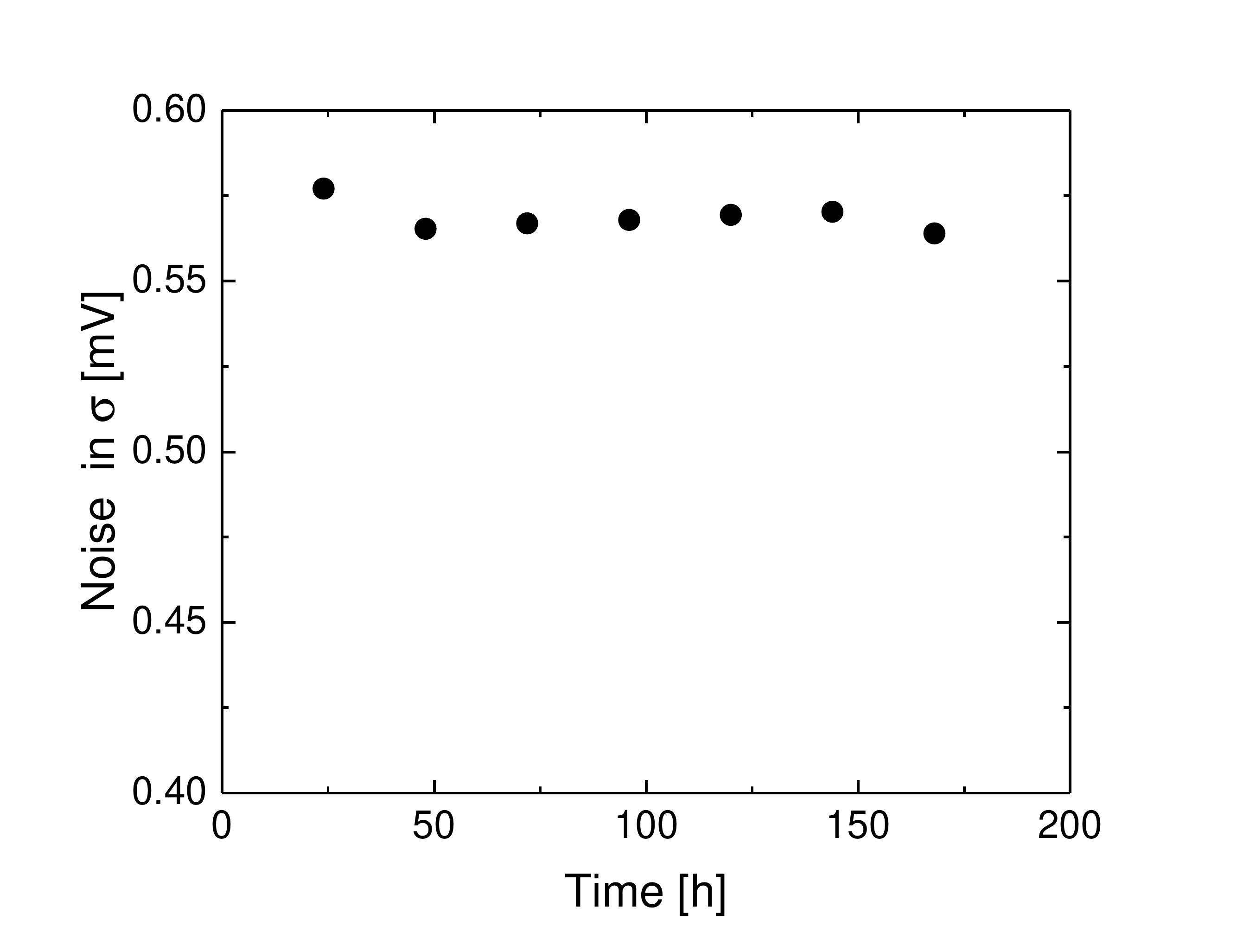}
\caption{Stability over time. Left: GCU temperatures; a stable temperature of
about $40^\circ\mbox{C}$ - $50^\circ\mbox{C}$ is reached after 16 hours
(about 1000 minutes) after power-on.
Right: Noise level as a function of time, monitored for one week.}
\label{fig:BX:tnvstime}
\end{figure}

The electronics is cased by an air-filled, stainless steel housing,
which in turn was glued to the neck of the PMT with several epoxies.
The glue joint and the cable feedthrough were covered with a heat-shrinkable
tube as a second layer of leakage protection. 
The performances were measured again after the potting procedure.
The temperature of the GCU was monitored with four temperature sensors
inside the FPGA. The temperature trend,
is shown in the left plot of Figure~\ref{fig:BX:tnvstime}.
After a fast initial
increase, the temperature stabilizes around $40^\circ\mbox{C}$ to
$50^\circ\mbox{C}$. There is no significantly change over the next
220 hours of measurement.
During the operation the outside air temperature was stabilized with
ventilation to $25^\circ\mbox{C}$.
A somewhat better cooling is expected in water.
Position and width of the baseline were stable during the whole period.
From the width we extract a noise level corresponding to 
$0.60 \pm 0.04~\mbox{mV}$.
It is shown in the right plot of Figure~\ref{fig:BX:tnvstime}
The signal-to-noise ratio has been measured to be
$18.0 \pm 1.6$, a value which is compatible, within the statistical error
with that obtained before sealing the PMT with the electronics inside.

\begin{figure}[htb]
\centering
\includegraphics[scale=0.4]{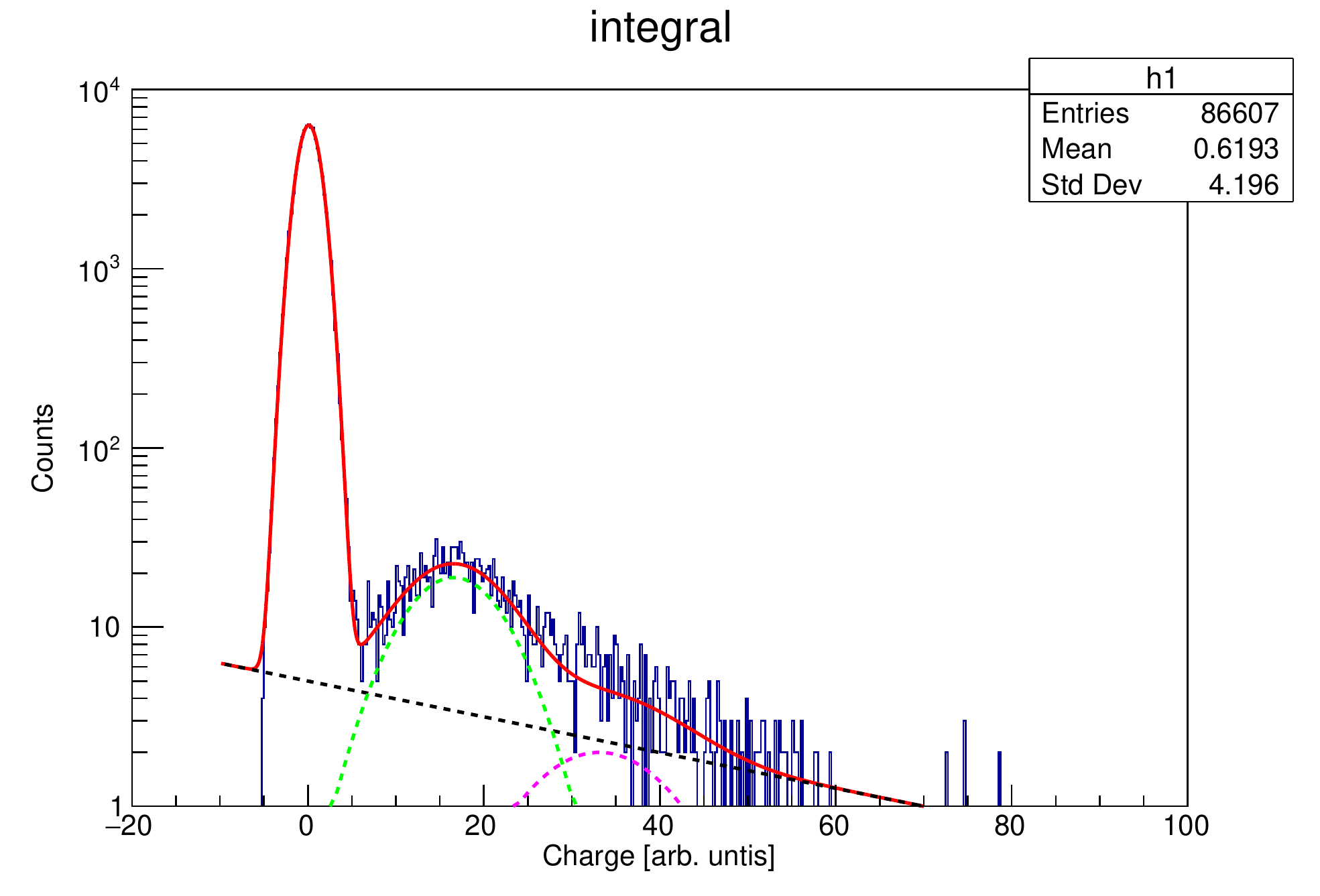}
\caption{Single p.e. spectrum, after potting.}
\label{fig:BX:single_pe:PtV:potting}
\end{figure}

We recorded single photon spectra with a pulsed LED in front of the photo
cathode. A trigger generated by the pulse generated was sent to the GCU.
The data was recorded through the full vertical slice. 
There was no visible change in the rise or decay time of the pulses
after potting. Again the pulses were integrated over 50~ns.
The charge spectrum is shown in Figure~\ref{fig:BX:single_pe:PtV:potting}.
The signal-to-noise ratio of the single p.e. signal, which was estimated at
about 10, did not significantly change either.
The fit of the charge spectrum is explained in
Figure~\ref{fig:BX:single_pe:PtV}.
The different contributions of the fit are reported in
Figure~\ref{fig:BX:single_pe:PtV}.
The single p.e. resolution is around $34 \%$, which is also compatible
to the results before the potting procedure.

\section{Conclusions}
\label{sec:conclusions}
JUNO will be the largest liquid scintillator detector ever
built for neutrino physics.
The scientific goals put stringent constraints on the performance of the
readout electronics. Especially challenging are the excellent energy resolution
required for the determination of the mass hierarchy, the large data rate from
supernova events due to the large mass of the detector and the handling of the
huge signals of cosmic muons.
The readout electronics of the large PMTs is an essential ingredient for the
success of the experiment. A novel design of the electronics
has been presented.
The electronics is mounted on the back end of the PMTs
to the PMT output signal, embedded in the watertight steel housing.
A substantial effort has gone into optimizing the reliability of the system.
The tests confirm the expected performance of the whole system.
It was verified that the potting does not degrade the performances.

\acknowledgments
Part of this work has been supported by the Italian-Chinese
collaborative research program jointly funded by the Italian Ministry of
Foreign Affairs and International Cooperation (MAECI) and the National
Natural Science Foundation of China (NSFC). We also acknowledge the
support by the Deutsche Forschungsgemeinschaft, DFG, FG 2319 and
of the F.R.S-FNRS funding agency (Belgium).


\end{document}